\documentclass[usenatbib]{mnras}
\usepackage{bookmark}
\usepackage{footmisc}  
\usepackage[usenames,dvipsnames]{xcolor}
\usepackage[flushleft]{threeparttable}
\usepackage{graphicx}	
\usepackage{epsfig}
\graphicspath{ {} {} }
\usepackage{multirow}
\usepackage{amsmath}	
\usepackage{amssymb}	
\usepackage{times}
\usepackage{newtxtext,newtxmath}
\usepackage[T1]{fontenc}
\usepackage{ae,aecompl}
\usepackage{tabularx}
\usepackage {enumerate} 
\usepackage[normalem]{ulem}

\usepackage{listings}
\PassOptionsToPackage{unicode}{hyperref}
\PassOptionsToPackage{naturalnames}{hyperref}

\newcommand{\hii}{H{\sc ii}}
\newcommand{\uchii}{UC\,H{\sc ii}}
\newcommand{\hchii}{HC\,H{\sc ii}}

\title[]{A search for hypercompact \hii\ regions in the Galactic Plane}

\author[A. Y. Yang et al.]{
A. Y. Yang,$^{1,2,3,6}$
M. A. Thompson,$^{3}$
W.W. Tian,$^{1,2,5}$
S. Bihr,$^{4}$
H. Beuther$^{4}$,
L. Hindson$^{3}$\\
$^{1}$Key Laboratory of Optical Astronomy, National Astronomical Observatories, Chinese Academy of Sciences, Beijing 100012, China\\
$^{2}$University of Chinese Academy of Sciences, 19A Yuquan Road, Beijing 100049, China\\
$^{3}$Centre for Astrophysics Research, School of Physics Astronomy \& Mathematics, University of Hertfordshire, College Lane, Hatfield, AL10 9AB, UK\\
$^{4}$Max-Planck Institute for Astronomy, K\"onigstuhl 17, 69117 Heidelberg, Germany\\
$^{5}$Department of Physics $\&$ Astronomy, University of Calgary,  Alberta t1N 1N4, Canada\\
$^{6}$Max Planck Institute for Radio Astronomy, Auf dem H\"ugel 69, 53121, Bonn, Germany\\
}

\date{Accepted 2018. Oct.14; Received 2018 Oct.; in original form 2017 October 13}

\pubyear{2017}
\begin{document}
\label{firstpage}
\pagerange{\pageref{firstpage}--\pageref{lastpage}}
\maketitle
\renewcommand{\thefootnote}{\fnsymbol{footnote}}

\begin{abstract}

We have carried out the largest and most unbiased search for hypercompact (HC) \hii\ regions. Our method combines four interferometric radio continuum surveys (THOR, CORNISH, MAGPIS and White2005) with far-infrared and sub-mm Galactic Plane surveys to identify embedded \hii\ regions with positive spectral indices. 120 positive spectrum \hii\ regions have been identified from a total sample of 534 positive spectral index radio sources. None of these \hii\ regions, including the known \hchii\ regions recovered in our search, fulfills the canonical definition of an \hchii\ region at 5 GHz. We suggest that the current canonical definition of \hchii\ regions is not accurate and should be revised to include a hierarchical structure of ionized gas that results in an extended morphology at 5 GHz. 
Correlating our search with known ultracompact (UC) \hii\ region surveys, we find that roughly half of detected \uchii\ regions have positive spectral indices, instead of  more commonly assumed flat and optically thin spectra. This implies a mix of optically thin and thick emission and has important implications for previous analyses which have so far assumed optically thin emission for these objects. Positive spectrum \hii\ regions are statistically more luminous and possess higher Lyman continuum fluxes than \hii\ regions with flat or negative indices. Positive spectrum \hii\ regions are thus more likely to be associated with more luminous and massive stars. No differences are found in clump mass, linear diameter or luminosity-to-mass ratio between positive spectrum and non-positive spectrum \hii\ regions.

\end{abstract}

\section{Introduction}
\label{sect:intro}

Newly formed massive stars are deeply embedded within molecular clouds, 
but they produce powerful Lyman continuum emission that is sufficiently energetic to ionize their surroundings and create  observable ionized regions, known as \hii\ regions. \hii\ regions are over-pressured with respect to the surrounding interstellar medium and so expand over time, driving ionization shocks into the ambient medium \citep{Tenorio1979AA71,Dyson1995MNRAS277}.  
The smallest \hii\ regions are thus likely to be the youngest, which means that the most dense and compact \hii\ regions can shed light on the early development of massive stars \citep{Peters2010ApJ711}.  

However, many details of these early stages are currently unclear \citep{Hoare2007prplconfH,Zinnecker2007ARAA45}, and different theoretical models predict very different outcomes for the earliest stages of \hii\ regions. For example, as massive stars reach the main sequence while still accreting material, the
\citet{McKee2003ApJM} and \citet{Peters2010ApJ711} turbulent core and ionization feedback models both envisage the youngest \hii\ regions expanding into outflow-driven cavities away from the main accretion flows \citep[e.g.][]{Tan20039139T,Tanaka2016ApJ52T}. This implies the early development of \hii\ regions within the star formation process. 
However, the models of \cite{Hosokawa2009ApJ691} and \cite{Hosokawa2010ApJ721}, suggest that the high accretion rates of material onto the massive star cause its outer layers to swell, reducing the stars effective temperature. This has the result of delaying the initial development of \hii\ regions until after the accretion phase has finished.

At first sight, the difference in the onset time for \hii\ regions from different models implies that the relative incidence of the smallest and earliest \hii\ regions may allow different models to be discriminated. However, subsequent differences in the expansion rate of the \hii\ regions may  complicate matters. For example, the rapid expansion of \hii\ regions into outflow cleared cavities \citep[e.g.][]{Tan20039139T,Tanaka2016ApJ52T} in the \citet{McKee2003ApJM} and \citet{Peters2010ApJ711} may result in a dearth of small \hii\ regions despite the early onset of the ionization. Nevertheless, it is clear that the main differences in these theoretical models arise at the earliest onset of the \hii\ region phase and that studying the physical properties of youngest and smallest HII regions will help to improve our theoretical understanding of massive star formation.

 Observationally, the smallest and densest \hii\ regions so far discovered are commonly known as hyper-compact \hii\ regions (\hchii) that have typical size $\rm \lesssim 0.03\,pc$, electron density $\rm \gtrsim 10^{6}\,cm^{-3}$, and emission measure $\rm \gtrsim 10^{10}\,pc\,cm^{-6}$ \citep{Kurtz2005IAUS}. 
 They are distinguished from the next largest class of \hii\ regions, the ultra-compact \hii\ regions (\uchii) that have typical size $\rm \lesssim 0.1\,pc$, electron density $\rm \gtrsim 10^{4}\,cm^{-3}$, and emission measure $\rm \gtrsim 10^{7}\,pc\,cm^{-6}$ \citep{Kurtz2005IAUS}, 
 primarily because they show extremely broad radio recombination lines (RRL) \citep{Hoare2007prplconfH}, 
with typical $\rm \Delta V = 40-50\,km\,s^{-1}$ and some regions displaying $\rm \Delta V \gtrsim 100\,km\,s^{-1}$ 
\citep{Gaume1995ApJ,Johnson1998ApJ,Sewilo2004ApJ}. By comparison, \uchii\ regions typically have $\rm \Delta V = 25-30 \,km\,s^{-1}$
\citep{Wood1989ApJS,Afflerbach1996ApJS106}. 

We summarize the observational properties of known \hchii\ regions from the literature in Table \ref{summary_hchii}. 
As can be seen, the RRL line widths of \hchii\ regions, their emission measures (EM) and electron densities ($n_{\rm e}$) are in general larger than those of \uchii\ regions, although there is a considerable spread of values. The radio spectral indices of \hchii\ regions are positive with a typical value $\sim $1 \citep{Beuther2007prpl165B}, i.e., between the purely optically thick and optically thin values discussed by \citet{Kurtz2005IAUS} which may, in turn, indicate line-of-sight density gradients \citep{Keto2003ApJ}.

Another property of \hchii\ regions is their rarity, 
with only 16 confirmed \hchii\ regions in the literature compared to 
$\sim$600 \uchii\ regions \citep{Urquhart2013MNRAS,Lumsden2013ApJS208,Cesaroni2015AA579A}. However, it is not clear if this rarity is intrinsic or caused by observational biases. As most surveys for young \hii\ regions have been carried out between 1.4 and 5 GHz, they unfortunately suffer from an observational bias towards the discovery of objects with flat or falling spectral indices. \hchii\ regions with strong positive spectra have 1.4--5 GHz fluxes orders of magnitude less than \uchii\ regions and are thus likely to have been missed in  low-frequency surveys. The majority of the \hchii\ regions listed in Table \ref{summary_hchii} were discovered serendipitously by  high-frequency observations of known \uchii\ regions. 

Our current limited understanding of the number and global properties of \hchii\ regions means that it is difficult to constrain theoretical models of 
massive star formation as these models predict very different evolutionary scenarios for \hii\ regions. Secondly, it is also difficult to place \hchii\ regions into context with the more well-known \uchii\ regions. For example, the defining characteristics of size, electron density, and emission measure proposed by \citet{Kurtz2005IAUS} fall on a continuous spectrum and so it is difficult to identify discrete ``types''. As we can see in Table \ref{summary_hchii} the sizes, emission measures, and electron densities of many of the identified ``\hchii\  regions'' fall within ranges that are more appropriate to \uchii\ regions. With only a handful of identified \hchii\ regions, it is difficult to identify a clear dividing line between the two types of \hii\ regions. Understanding the evolution of the youngest and densest \hii\ regions requires more confirmed examples to be found.

We have undertaken the largest and most systematic search yet carried out for \hchii\ regions. As Table \ref{summary_hchii} shows, the easiest defining observational characteristic of \hchii\ regions is their positive radio spectral index. While broad recombination line widths are a key observational feature of \hchii\ regions, routine spectroscopy of large numbers of faint Galactic radio sources will not be possible until the advent of the SKA \citep[e.g.][]{thompson2015}.  We have used current radio interferometer surveys of the Galactic Plane, namely CORNISH \citep[5\,GHz,][]{Hoare2012PASP}, MAGPIS \citep[1.4\,GHz,][]{Helfand2006AJ,White2005AJ}, 
and THOR \citep[1$\sim$2\,GHz,][]{Bihr2016AA,Beuther2016AA595A32B}, to identify objects with positive spectral indices between 1.4 and 5 GHz. We then cross-match these positive spectrum radio sources with mid/far-infrared and submillimetre survey data to identify objects embedded within molecular clumps, as would be expected for the initial stages of massive star formation. This is very similar to the strategy that was used to classify sources in CORNISH\footnote{ http://cornish.leeds.ac.uk/public/cone.php \label{third_footnote}}. Of course, the sample that we have assembled here is also subject to the observational bias towards negative and flat spectrum indices that we have previously mentioned. However, we can nevertheless identify bright \hchii\ regions within the CORNISH catalogue and obtain a much wider understanding of young \hii\ regions.

This paper is organized as follows. Section 2 describes the Galactic Plane surveys we used to identify positive spectrum radio objects and confirm those that are embedded within molecular cloud clumps. We describe the procedure used to determine spectral indices and identify 
a sample of objects with rising radio spectra in Section 3. 
In Section 4, we characterize the properties of the sample at  far-infrared (FIR) and submillimeter (sub-mm) wavelengths, select a subsample of young and compact \hii\ regions and discuss their properties. In particular, we compare the properties of \hii\ regions with rising (positive) radio spectra to those with flat or negative spectra and dwell upon the implications of our survey for the frequency of \hchii\ regions in the Galaxy. In Section 5 we present a summary of our conclusions.

\begin{table*}
\setlength{\tabcolsep}{0.005pt}
\scriptsize
 \caption{\large Summary of \hchii\ regions: Name and Galactic name, radio recombination lines (RRL) and its line width (FWHM), spectral indices, heliocentric distance (Dist.), linear diameter (Diam.), electron density ($n_{e}$), emission measure (EM), and reference.}
  \begin{tabular}{p{3cm}p{3cm}p{2cm}p{3cm}p{1.3cm}p{1.3cm}p{1.3cm}p{1.3cm}p{1.3cm}}
  \hline
  \hline
  Name       & Gname   & RRL\,(FWHM)  & Spectral Index & Dist.& Diam. & $\rm n_{e}$ & EM &  Reference \\
 & & Hn$\alpha$\,(km\,$s^{-1}$)    & & (kpc) & (pc) & ($\rm 10^{5}\,cm^{-3})$& ($\rm 10^{8}\,pc\,cm^{-6}$)   & \\
  \hline
  Sgr B2 F & G000.6667$-$00.0362  & H66$\alpha$\,(80)& 0.95\,(1.4-22.5GHz) & 8.5 & 0.011 & $\sim 10$ & $\sim 10$  & 3,6 \\
G10.96$+$0.01 W &  G010.9583$+$00.0223  & H92$\alpha$\,(43.8) & $\sim$1.2\,(1.4-5GHz) & 14 &0.121 & 0.29 & 0.53 & 6,10\\
M17-UC1 & G015.0346$-$00.6771  & H66$\alpha$\,(47) &  1.1\,(1.5-22GHz) &  2.2 & 0.006 & 3.3 & 2.6 &5,6 \\
G24.78$+$0.08A1 & G024.7898$+$00.0833 & H66$\alpha$\,(40)&  - &  7.7 & 0.002  &  $-$ & $-$ &  8  \\  
G28.2$-$0.04 N  &  G028.2002$-$00.0495 & H92$\alpha$\,(74) & 1.0$\pm0.1$\,(1.4-15GHz) & 5.7 & 0.028 & 0.76 & 1.9 & 6,10,12     \\ 
G34.26$+$0.15 B & G034.2580$+$00.1533 & H76$\alpha$\,(48.4) & $0.9\pm0.4$(5-23GHz) & 3.3 & 0.008 & 2.2 & 4.3 &  6,7  \\
G35.58$-$0.03 & G035.5780$-$00.0313 & H30$\alpha$  \,(43.2) & 0.32$\pm$0.04\,(8.3-23GHz) & 10.2 & 0.018 & 3.3 & 19 &  9\\
W49 AA & G043.1653$+$00.0128 & H66$\alpha$\,(53.7) & 0.6\,(8.3-43GHz) & 11.4 & 0.035 & 0.55 & 2.2 & 4,6 \\
W49 AB  & G043.1660$+$00.0120 & H66$\alpha$\,(63.9) & $\sim$1.1(8.3-43GHz) & 11.4 & 0.031 & 0.45 & 1.3 & 4,6 \\
W49 AG  & G043.1666$+$00.0110  & H66$\alpha$\,(48.6) & $\sim$2\,(22-43GHz) & 11.4 & 0.061 & 0.33 & 1.4 & 4,6 \\
G45.07$+$0.13 NE & G045.0712$+$00.1322 & H76$\alpha$\,(40) & $-$ & 6.0 & 0.032 & 0.94 & 1.9   & 10,12\\
$\rm W51e2^{a}$  & G049.4898$-$00.3874    & H66$\alpha$\,(54) & $-$ & 7.0 & 0.02   & 16 & $-$ & 12,13 \\
NGC 7538-IRS1 & G111.5382$+$00.8112 & H66$\alpha$ \,(180) & $\sim$0.9\,(23-50GHz)  & 3.5 & 0.04 & 1.2 & 2.1 & 1,6,12      \\
G301.1366$-$00.2248  & G301.1366$-$00.2248 & H70$\alpha$\,(66) & 1.5\,(0.843-20GHz) &  4.5 & 0.02 &  6.8 & 30 & 11  \\
G309.9217$+$00.4788  &  G309.9217$+$00.4788 & H70$\alpha$\,(40) & 1.2\,(0.843-20GHz) &  5.5 & 0.03 & 2.3 & 8 & 11 \\
G323.4594$-$00.0788  & G323.4594$-$00.0788 & H70$\alpha$\,(50) & 0.8\,(0.843-20GHz) & 4.8\,/\,8.9 & 0.05 & 0.64 & 1.9 & 11 \\
\hline
\end{tabular}
\begin{tablenotes}  
\item 
Reference: 1, \cite{Gaume1995ApJ}; 2, \cite{Shepherd1995ApJ}; 
3, \cite{dePree1996ApJ}; 4, \cite{dePree1997ApJ,DePree2004ApJ}; 5, \cite{Johnson1998ApJ}; 6, \cite{Sewilo2004ApJ}. 7, \cite{Avalos2006ApJ}; 8, \cite{Beltran2007AA}; 9, \cite{Zhang2014ApJ}; 10, \cite{Sewilo2011ApJS}; 11, \cite{Murphy2010MNRASa2}; 12, \cite{Keto2008ApJ672}; 13, \cite{Shi2010ApJ843S}.
\end{tablenotes}
\label{summary_hchii}
\end{table*}    
\begin{table*}
\setlength{\tabcolsep}{0.005pt}
\scriptsize
\caption {\large Galactic Plane surveys used in this study: survey name, wavelength and beam size, $1\sigma$ sensitivity, longitude ($\ell$) and latitude ($b$) coverage, and reference.}   
\begin{tabular}{p{3cm}p{3cm}p{3cm}p{4.5cm}p{2.5cm}p{1.5cm}}
\hline
\hline
    Survey &  Wavelengths $\&$Beam & Sensitivity   &   $ \ell$ coverage                & $ b$ coverage    &  Reference    \\
      & &  (mJy beam$^{-1}$)   &   $\degr$ & $\degr$    &      \\
 \hline
   CORNISH &  5\,$\rm GHz $ $\&$ 1.5$\arcsec$   & 0.4 (1$\sigma$) &  10$<$ $l$ $<$ 65\degr &$\mid b\mid$ $<$ 1\degr &  1 \\
   \hline
   MAGPIS	 &  20$\,\rm cm $ $\&$ 6$^{''}$	&$\sim$0.3 (1$\sigma$)      & 10$<$l$<$32\degr &  $\mid b\mid$ $<$ 0.8\degr &   2\\
   \hline
   THOR &  1$-$2\,$\rm GHz$ $\&$ 10$-$25$\arcsec$ & 0.3$-$1 (1$\sigma$) &   14.0$<\,$l$\,<$37.9\degr $\&$ 47.1<\,$l$\,<51.2\degr  &   $\mid b\mid$ $\leq$ 1\degr & 3 \\
\hline

\multirow{2}{*}{White2005} &  \multirow{2}{*}{20\,$\rm cm$ $\&$ 6$^{''}$}& \multirow{2}{*}{$\sim$0.9 (1$\sigma$)}    & 10$<$ $l$ $<$ 39\degr &   $\mid b\mid$ $<$ 1\degr &  \multirow{2}{*}{4}  \\
 & & & 39$\leq$ $l$ $<$ 65\degr &   $\mid b\mid$ $<$ 0.8\degr                          \\
\hline
  \multirow{2}{*}{\footnotesize Total overlapping sky region}&      & $l$ = 10$-$39\degr ,47.1$-$51.2\degr &$\&$ & \hspace{-30mm} $\mid b\mid $ $<$ 1\degr      &      \\
 &   & $l$ = 39.0$-$65\degr &  $\&$ & \hspace{-30mm} $\mid b\mid $ $<$ 0.8\degr &             \\
\hline
   Hi-GAL        & 250\,$\mu m$ $\&$ 18\arcsec  & $\sim$ 12.8\,(1$\sigma$)          &  $\mid l \mid$ $\leq$ 60\degr &$\mid b\mid$ $\leq$ 1\degr & 5 \\
 \hline

   ATLASGAL & $870\,\mu m$ $\&$ 19\arcsec	&$\sim$50$-$70\,(1$\sigma$)      & -60$<$ $l$ $<$ 60\degr   &  $\mid b\mid$ $<$ 1.5\degr    &   6   \\
   \hline
   \hline 
\end{tabular}
\begin{tablenotes}  
\item
Reference: 1, \cite{Hoare2012PASP,Purcell2013ApJS}; 2, \cite{Helfand2006AJ}; 
3, \cite{Bihr2016AA,Beuther2016AA595A32B}; 4, \cite{White2005AJ}; 
5, \cite{Molinari2010PASP}; 6, \cite{Schuller2009AA}.
\end{tablenotes}
\label{survey_info}
\end{table*}    

\begin{table*}
\setlength{\tabcolsep}{0.005pt}
\scriptsize
\caption {\large Cross-matching procedure}   
\begin{tabular}{p{3.5cm}p{3.5cm}p{3.5cm}p{3.5cm}p{4cm}}
\hline
\hline
 \multicolumn{5}{c}{\footnotesize Selection process of 1.4 GHz counterparts with $\alpha-d\alpha > 0$ }     \\

   \hline
 Overlapping CORNISH & Match selection & Unique matched sources & Positive spectrum objects & unresolved (resolved) \\
  \hline
THOR  &     matching radius $=$ 20\arcsec &    remove duplicates & $\alpha-d\alpha > 0.0$  & $S_{int} \leq 1.2S_{peak}$ ($S_{int} > 1.2S_{peak}$) \\
 \hline
 $l$ = 14.0$-$37.9\degr,  $\mid b\mid $ $<$ 1\degr & \multirow{2}{*}{1060\hspace{8mm}$\rightarrow $}  & \multirow{2}{*}{556\, \hspace{8mm}$\rightarrow $}     & \multirow{2}{*}{108\,\hspace{8mm}$\rightarrow$}  & \multirow{2}{*}{106\,(2)} \\
 $l$ = 47.1$-$51.2\degr,  $\mid b\mid $ $<$ 1\degr&   & &  &   \\
\hline
    MAPGIS &  matching radius $=$ 6\arcsec & remove duplicates & $\alpha-d\alpha > 0.0$  & $S_{int} \leq 1.2S_{peak}$ ($S_{int} > 1.2S_{peak}$) \\
 \hline
 $l$ = 10$-$32\degr, $\mid b\mid$ $<$ 0.8\degr &647\,\hspace{8mm}$\rightarrow $  & 647\,\hspace{8mm}$\rightarrow $ & 126\,\hspace{8mm}$\rightarrow$ & 58 (68)\\
$l$ = 32$-$48.5\degr,  $\mid b\mid $ $<$ 0.8\degr & 53\,\,\hspace{8mm}$\rightarrow$ & 53\,\,\,\hspace{8mm}$\rightarrow$ & 25\,\,\,\hspace{8mm}$\rightarrow$ & 24 (1)  \\
 \hline
  White2005 &  matching radius $=$ 6\arcsec & remove duplicates & $\alpha-d\alpha > 0.0$  & $S_{int} \leq 1.2S_{peak}$ ($S_{int} > 1.2S_{peak}$) \\ 
\hline
 10$<$ $l$ $<$ 39\degr, $\mid b\mid$ $<$ 1\degr &  \multirow{2}{*}{963\,\hspace{8mm}$\rightarrow $} & \multirow{2}{*}{397\,\hspace{8mm}$\rightarrow $} & \multirow{2}{*}{151\,\hspace{8mm}$\rightarrow$} & \multirow{2}{*}{68\,(83)}\\
 39$\leq$ $l$ $<$ 65\degr, $\mid b\mid$ $<$ 0.8\degr         &   &          &  & \\
 \hline
\footnotesize In total &2723\hspace{8mm}$\rightarrow $  & 1653\hspace{8mm}$\rightarrow $ & 410\,\hspace{8mm}$\rightarrow$  & 256 (154)\\
 \hline
  \multicolumn{5}{c}{\footnotesize Selection process of 1.4\,GHz dropouts with $\alpha_{min} > 0$  }     \\
   \hline
 Overlapping CORNISH& Unmatched selection   & Exclude extended source & Unique sources & Positive spectrum objects  \\
  \hline
   THOR & Removing those within  30\arcsec & $d_{\rm sep}$ > $1.2d_{\rm THOR\,source}$ &  remove duplicates & ${\alpha_{\rm min}} > 0.0$ \\
 \hline
 $l$ = 14.0$-$37.9\degr,  $\mid b\mid $ $<$ 1\degr & \multirow{2}{*}{173\,\,\hspace{8mm}$\rightarrow $}  & \multirow{2}{*}{123\,\hspace{8mm}$\rightarrow $}   & \multirow{2}{*}{118\,\hspace{8mm}$\rightarrow$}  & \multirow{2}{*}{87}     \\
 $l$ = 47.1$-$51.2\degr, $\mid b\mid $ $<$ 1\degr &   & &  &   \\
\hline
   MAPGIS        & Removing those within 10\arcsec & $d_{\rm sep}$ > $1.2d_{\rm MAGPIS\,source}$     &     remove duplicates   &${\alpha_{\rm min}} > 0.0$  \\
 \hline
 $l$ = 10$-$32\degr, $\mid b\mid$ $<$ 0.8\degr &102\,\hspace{8mm} $\rightarrow $  & 101\,\hspace{8mm}$\rightarrow $ & 6\,\,\,\hspace{8mm}$\rightarrow $  & 6 \\
\hline
 White2005 & Removing those within 10\arcsec & $d_{\rm sep}$ > $1.2d_{\rm White2005\,source}$ & remove duplicates  &  ${\alpha_{\rm min}} > 0.0$  \\ 
\hline
 10$<$ $l$ $<$ 39\degr, $\mid b\mid$ $<$ 1\degr &755\,\,\hspace{8mm}$\rightarrow $  &   716\,\hspace{8mm}$\rightarrow $ & 11\,\,\hspace{8mm} $\rightarrow $  &5 \\
 39$\leq$ $l$ $<$ 65\degr, $\mid b\mid$ $<$ 0.8\degr &566\,\hspace{8mm} $\rightarrow $  &   523\,\hspace{8mm} $\rightarrow $ &       53\,\,\hspace{8mm} $\rightarrow $  & 26 \\
 \hline
 \footnotesize In total &1569\hspace{8mm}$\rightarrow $  &   1463\hspace{8mm}$\rightarrow $ & 188\, \hspace{8mm}$\rightarrow $  &124 \\
\hline
\hline
\end{tabular}
\label{cross_match_info}
\begin{tablenotes}   
\item  Detailed description of our selection process are shown in section\,\ref{sect:matches} and section\,\ref{sect:method}. MAGPIS has only published catalog within the region $l$ = 5$-$32$^{\circ}$, $\mid b\mid$ $<$ 0.8$^{\circ}$ \citep{Helfand2006AJ}. For the remaining regions within the MAGPIS survey we use the AEGEAN photometry package \citep{Hancock2012MNRAS} to construct a source catalogue from image cutouts obtained from the MAPGIS website. 
Only a catalog within $l$ = 14.0$-$37.9$^{\circ}$, $l$ = 47.1$-$51.2$^{\circ}$, $\mid b\mid$ $\leq$ 1$^{\circ}$ has been published from THOR \citep{Bihr2016AA}.
Only the overlapping region with CORNISH is presented, and the total sky region of \cite{White2005AJ} is $l$ = -20$-$120$^{\circ}$ and  $b$ = $\pm$ 0.8$-$$\pm$2.7$^{\circ}$. 
\end{tablenotes}
\end{table*}    
\section{Galactic Plane surveys in the Radio, FIR, and Submm} 

In this section, we describe the individual Galactic Plane surveys that we have used to firstly identify a sample of positive spectrum radio sources, and secondly to determine that these sources are indeed embedded \hii\ regions.  

Our starting point was the CORNISH (Coordinated Radio ``N'' Infrared Survey for High-mass star formation), which we used to form a base 5 GHz radio source catalog. CORNISH was a sensitive ($\rm \sim 0.4 mJy\,beam^{-1}$) 
and high-resolution  ($\sim 1.5\arcsec$) 5 GHz survey of a section of the Galactic Plane (10$<$ $\ell$ $<$ 65\degr and $\mid b\mid$ $<$ 1\degr) , 
using the JVLA in B and BnA configuration. CORNISH detected 3062 continuum sources greater than 7$\sigma$ \citep{Hoare2012PASP,Purcell2013ApJS}. Together with CORNISH, we used three other 1.4 GHz radio surveys to determine 1.4--5 GHz spectral indices (or lower limits to spectral index where only 1.4 GHz upper limits were available), MAGPIS, THOR and the \citet{White2005AJ} VLA Galactic Plane survey.

The Multi-Array Galactic Plane Imaging Survey (MAGPIS)\footnote{ http://third.ucllnl.org/gps/index.html \label{second_footnote}} 
has the highest sensitivity and resolution of the 1.4 GHz surveys that we have used, with a resolution of $\sim 6\arcsec$ and noise level $\rm \sim 0.3\,mJy\,beam^{-1}$ \citep[][]{Helfand2006AJ}. 
A catalog has been published for the survey region 5$<$ $\ell$ $<$ 32\degr and $\mid b\mid$ $<$ 0.8\degr.  MAGPIS images are also available between $l$ = 32$-$48.5\degr and $\mid b\mid$ $<$ 0.8\degr, but no catalog of objects detected within these images has yet been published. We used image cutouts available from the MAPGIS website\footref{second_footnote} to identify 1.4 GHz counterparts to CORNISH sources in the uncatalogued region of the Galactic Plane. Further details of this process are described in Section \ref{sect:method}.

To cover the remainder of the  CORNISH survey region, we have used 1--2 GHz catalogs and data from The HI, OH, Recombination line survey of the Milky Way \citep[THOR,][]{Bihr2016AA,Beuther2016AA595A32B} 
and the 1.4 GHz VLA survey \citet[]{White2005AJ}, hereafter White2005. Details of all four surveys are summarised in Table~$\ref{survey_info}$. The complete region covered by all three 1.4 GHz surveys 
is $\ell$ = 10$-$39\degr $\&$ $\ell$ = 47.1$-$51.2\degr and $\mid b\mid $ $<$ 1\degr, 
$\ell$ = 39.0$-$65\degr and $\mid b\mid $ $<$ 0.8\degr. 

Once the 1.4--5 GHz radio spectral index of the objects has been derived, we select those objects with a positive spectral index and confirm that they are embedded within molecular cloud clumps by examining cutout images from the ATLASGAL and Hi-GAL surveys. The APEX Telescope Large Area Survey of the Galaxy (ATLASGAL), 
has a resolution of 19\arcsec\ and typical noise level of 50 to 70\,mJy beam$^{-1}$ \citep{Schuller2009AA}, 
providing the largest, unbiased database for detailed studies of large numbers of early stages of massive-star forming clumps in the Galaxy. 
 The Herschel infrared Galactic Plane Survey (Hi$-$GAL) imaged the Galactic Plane in five far-infrared and sub-millimetre bands, with the principal aim of detecting
the earliest phases of the formation of molecular clouds and high-mass stars \citep{Molinari2010PASP,Molinari2016AA591A}. 
We used the most sensitive band of Hi-GAL, i.e. the 250 $\mu$m band which has a resolution of 18\arcsec\ (similar to the ATLASGAL resolution of 19\arcsec) and 
  a 1$\sigma$ flux density sensitivity of 12.8\,mJy beam$^{-1}$. 
The 250 $\mu$m band lies close to the peak of the SED of cold dust and is thus an excellent tracer of molecular clumps surrounding embedded \hii\ regions.



\section{Determining the radio spectral index}
\label{sect:alpha}

The spectral index of a radio source is defined by a power law  relationship\footnote{note that in some, mainly extragalactic, studies Equation (1) is defined as $S\,  {\propto}\,  {\nu}^{-\alpha}$.}. between 
its flux density $S_{\nu}$ and frequency $\nu$ ($S_{\nu}\,  {\propto}\,  {\nu}^{+\alpha} $) 
 
In practice the spectral index is determined between two flux densities $S_{\nu_{1}}$  and $S_{\nu_{2}}$ 
measured at two specific frequencies $\nu_{1}$ and $\nu_{2}$ using the following equation

\begin{equation}\label{2}
{\alpha}=\frac {\log{(S_{\nu_{2}}/S_{\nu_{1}})}} { \log{(\nu_{2}}/{\nu_{1}})} \
\end{equation}

The true value of the spectral index $\alpha$ can only be determined when both fluxes in Equation \ref{2} are available. 
Nevertheless, it is possible to determine a lower limit to the spectral index $\alpha_{\rm min}$  for sources 
that are only detected at a higher frequency (5 GHz in our case) but not at the lower (1.4 GHz) by using an upper limit for the flux density at the lower frequency, i.e.

\begin{equation}\label{alpha_min}
{\alpha_{\rm min}}=\frac {\log{[(S_{5GHz}-dS_{5GHz})/(5*dS_{1.4GHz})]}} { \log{(5/1.4)} }\
\end{equation}

where $dS_{\nu}$ represents the 1$\sigma$ r.m.s.~error in the flux density at frequency $\nu$. 
Note that this equation assumes a 5$\sigma$ detection threshold to determine the upper limit and also subtracts the 1$\sigma$ error from the 5\,GHz measurement to determine a  reliable lower limit to the spectral index. 
In the following subsections we present our method of determining $\alpha$ and $\alpha_{\rm min}$ from the CORNISH, MAGPIS, THOR and White2005 surveys.
The individual steps in the process are summarized in Table \ref{cross_match_info}. 

\subsection{CORNISH sources with 1.4 GHz counterparts}

\label{sect:matches}

Using the CORNISH and MAGPIS/White2005/THOR source catalogs, 
we first identify CORNISH catalog sources that are positionally associated with corresponding MAGPIS/White2005/THOR catalog sources.
As the MAGPIS  and White2005 1.4 GHz surveys have resolutions of 6\arcsec, we use a circular matching threshold of 6\arcsec, however as THOR has a lower angular resolution of  10--25\arcsec, we relax the  matching threshold to 20\arcsec\, as used in THOR \citep{Bihr2016AA}.  
Using these matching criteria we obtain 700, 963, and 1060 matches to CORNISH sources from MAGPIS, White2005, and THOR respectively. 

Next, we remove duplicates caused by overlapping survey regions and create a unique list of matching sources by merging the MAGPIS, White2005 and THOR matches. Duplicates are removed by choosing the counterpart from the survey with the highest sensitivity at the CORNISH source position. In general, MAGPIS has the highest sensitivity, 
followed by THOR and finally by White2005. We must also take into account the fact that a MAGPIS catalog has not been published for the region between $\ell$ = 32$-$48.5\degr, although images are available from the MAGPIS project website. So, for all CORNISH sources lying in this region of the Plane, we obtained and inspected MAGPIS cutout images. For those CORNISH sources with 1.4 GHz counterparts, we measured their 1.4 GHz peak and integrated fluxes using the AEGEAN package \citep{Hancock2012MNRAS}. 53 CORNISH sources in this region were found to have MAGPIS counterparts. Overall, the matching and deduplication process resulted in 700 MAGPIS matches, 556 THOR matches and 397 White2005 matches respectively. 

Finally, as we are principally interested in positive spectrum sources in this study, 
we calculate the 1.4--5 GHz spectral index of each source using Equation \ref{2} and remove all sources whose spectral index $\alpha$ is less than zero. 
This results in a final positive spectrum (i.e.~$\rm \alpha-d_{\alpha}>0$) catalog of 410 sources; 
108 from THOR matches, 151 from MAGPIS matches, and 151 from White2005 matches. 

As this sample includes both point sources and extended sources, 
we need to bear in mind that the radio surveys are interferometric and carried out at different frequencies and with different VLA configurations. Each survey has a different coverage of the $uv$ plane, and thus we are only able to derive reliable spectral indices for point sources. 
We use the same criterion to distinguish between point and extended sources as the THOR survey \citep[][]{Bihr2016AA}, where extended sources are identified as those having an  integrated 1.4 GHz flux density more than 1.2 times their peak 1.4 GHz flux density ($S_{int} /S_{peak}>1.2$). 
Because lower frequency surveys are more sensitive to extended emission than higher frequency survey, 
the flux densities of extended objects at 1.4\,GHz surveys could contain more extended emission than 5\,GHz CORNISH. 
One way to address this would be to combine multiple VLA configurations with single-dish data to improve the $uv$ coverage and include zero-spacing flux \citep[e.g.][]{Tian2005AAT,Tian2006AAT}. However, this is not feasible for the archival survey data that we use here. 

\begin{figure*}
 \centering
  \begin{tabular}{cc}
    \includegraphics[width = 0.45\textwidth]{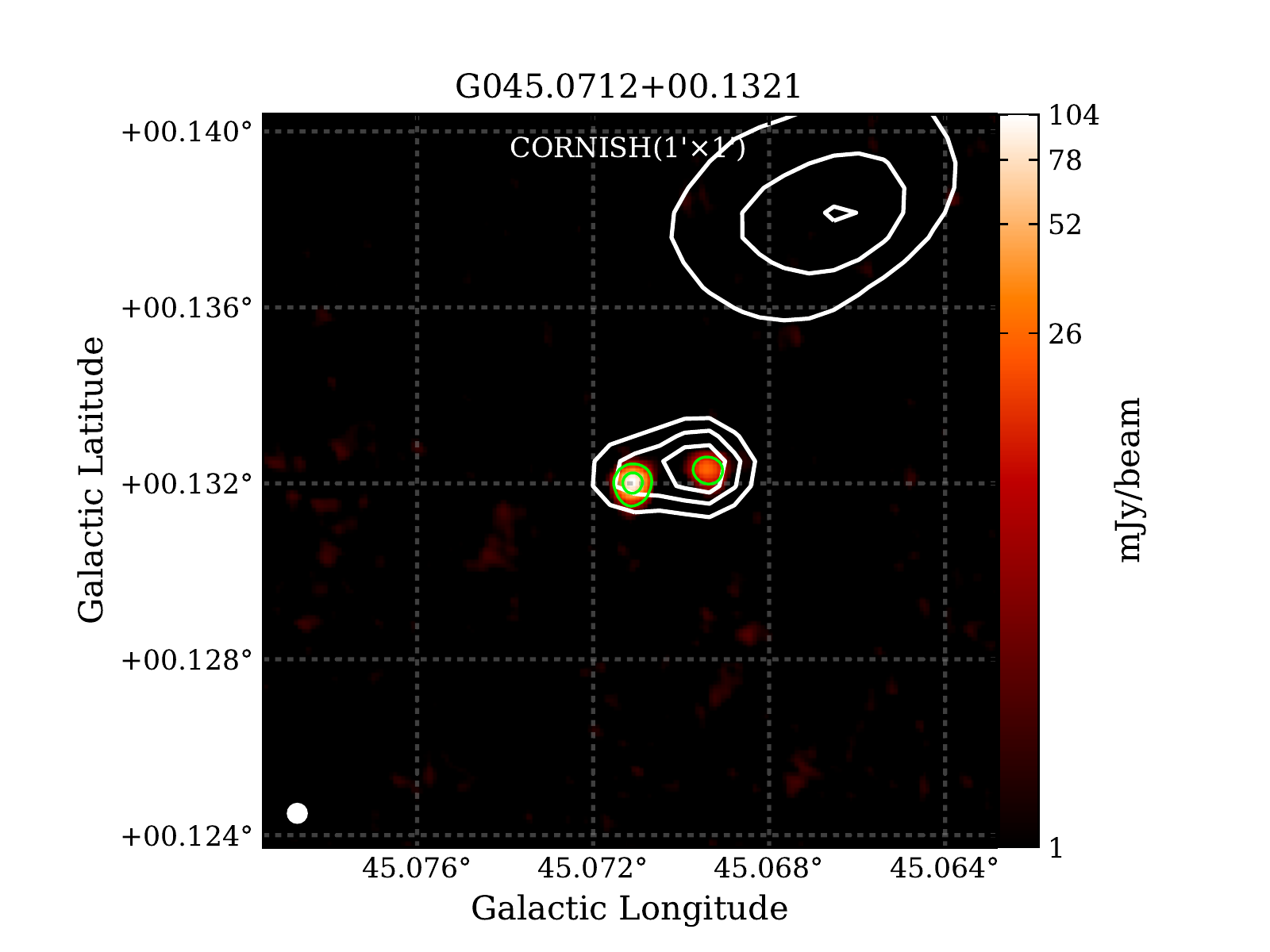}&\includegraphics[width = 0.45\textwidth]{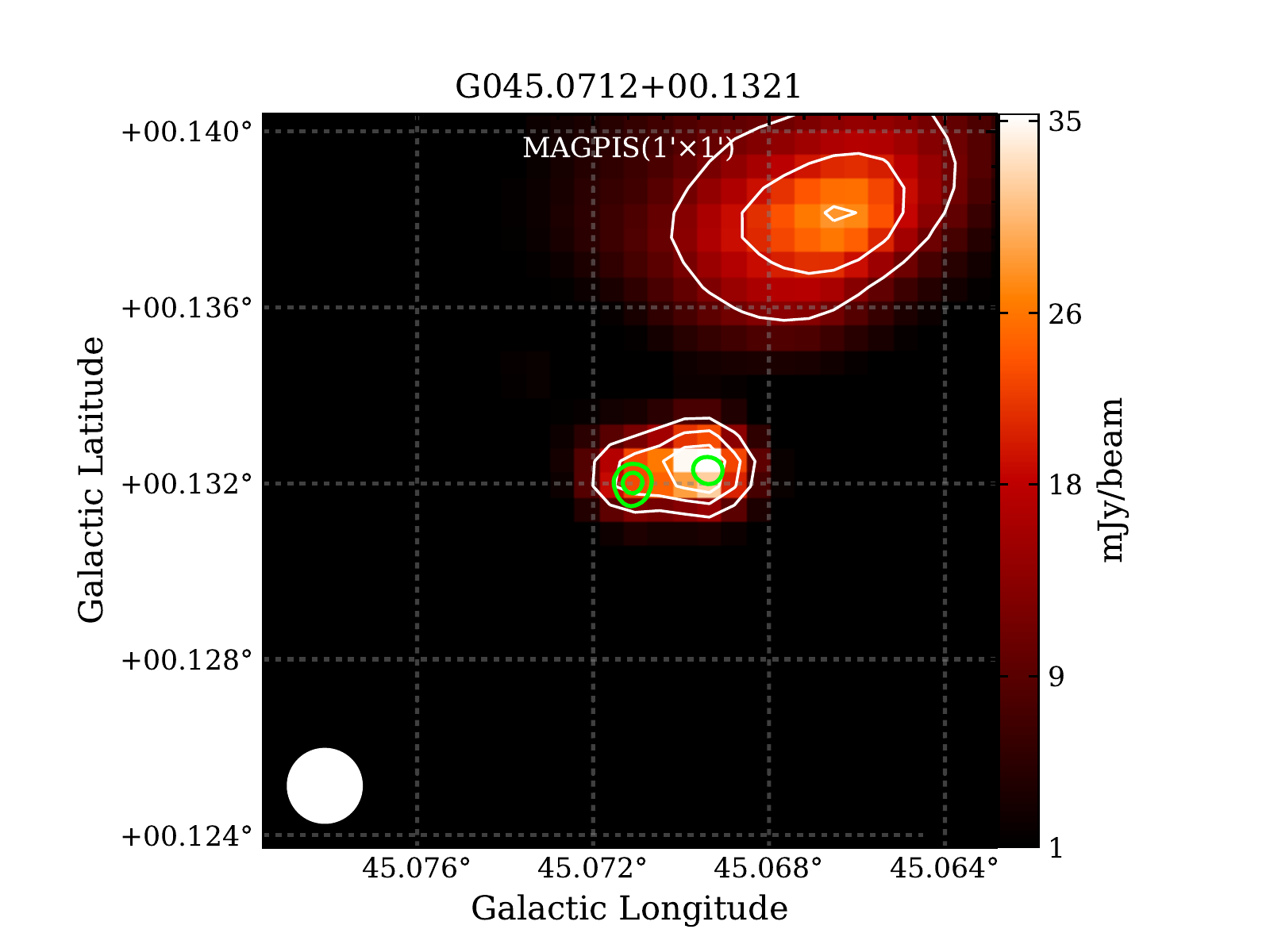}\\
  \end{tabular}
 \caption{Left: the image of known \hchii\ G45.07$+$0.13 NE (CORNISH counterpart: G045.0712+00.1321) at 5\,{\rm GHz} CORNISH. 
 Right: the image of known \hchii\ G45.07$+$0.13 NE at 1.4\,{\rm GHz} MAGPIS.  The beams of CORNISH (1.5\arcsec) and MAGPIS (6\arcsec) 
 are white circles presented in the lower left corner of each image. 
Overlaid lime and white contours on each image show 5\,GHz emission from CORNISH and 1.4\,GHz emission from MAGPIS, respectively. This is an example of an \hchii\ region showing extended 1.4 GHz emission  ($S_{int} /S_{peak}=1.74$) from a nearby \uchii\ region and we regard its spectral index as a strict lower limit ($\alpha_{min}$). }
 \label{radio_image_hchii_1}
 \end{figure*}

\begin{figure*}
 \centering
  \begin{tabular}{cc}
    \includegraphics[width = 0.45\textwidth]{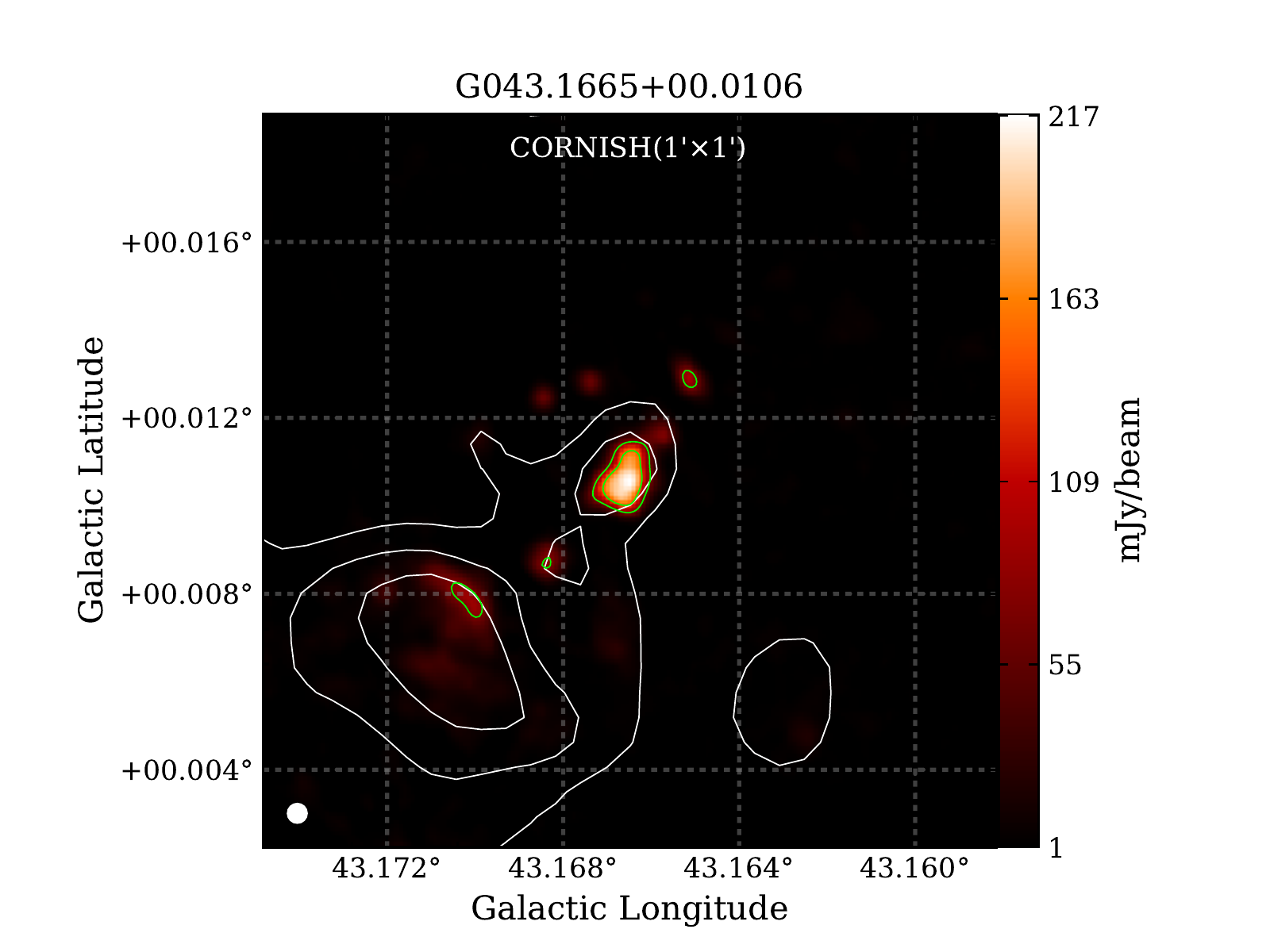}&\includegraphics[width = 0.45\textwidth]{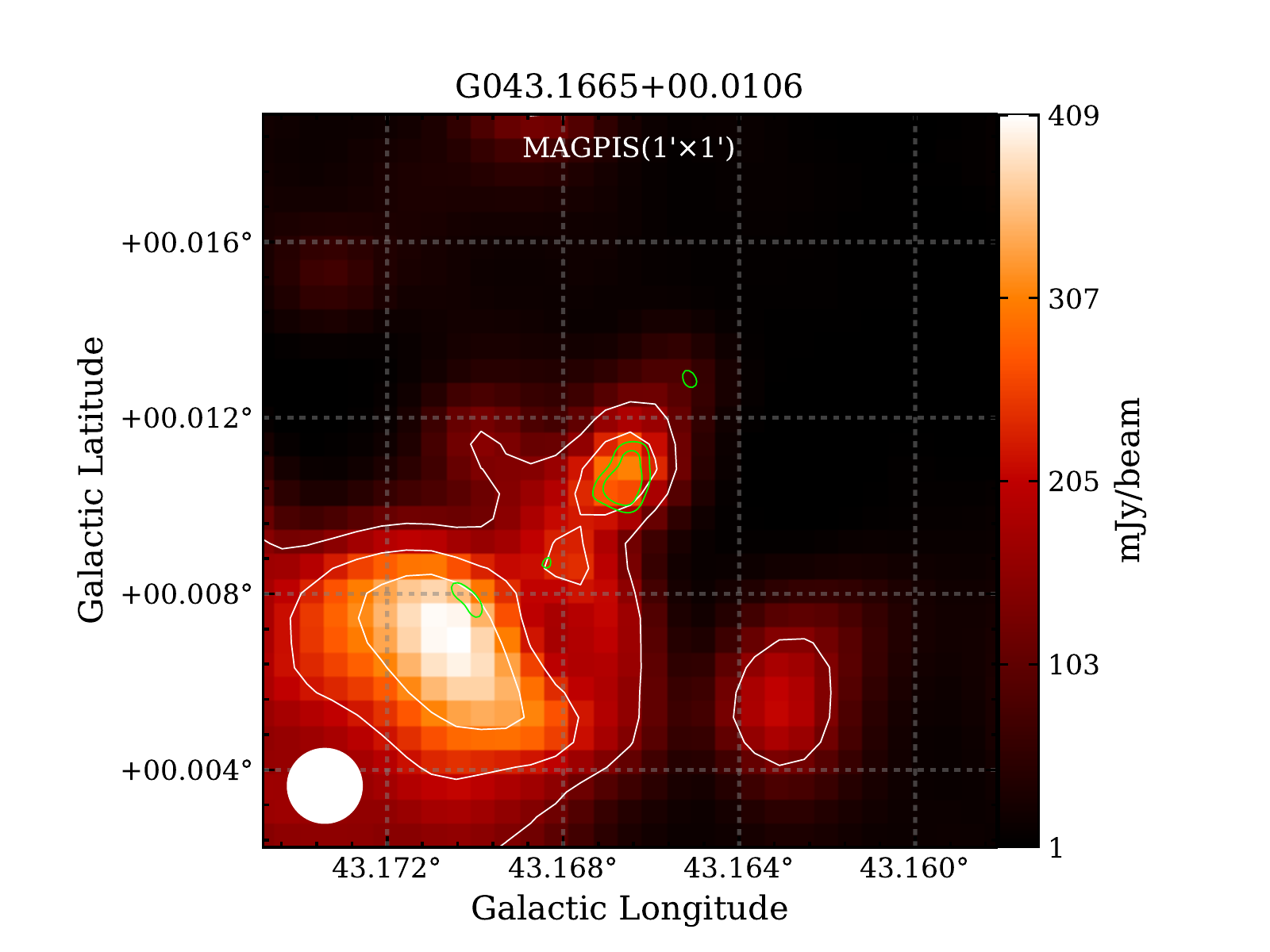}\\
  \end{tabular}
 \caption{ Left: the image of known \hchii\ W49A G (CORNISH counterpart: G43.1665$+$00.0106) at 5\,{\rm GHz} CORNISH. 
 Right: the image of known \hchii\ G43.1665$+$00.0106 at 1.4\,{\rm GHz} MAGPIS. The beams of CORNISH (1.5\arcsec) and MAGPIS (6\arcsec) are white circles presented in the lower left corner of each image. Overlaid lime and white contours on each image show 5\,GHz emission from CORNISH and 1.4\,GHz emission from MAGPIS, respectively. 
 This is an example of an \hchii\ region showing extended 1.4 GHz emission ($\rm S_{int} /S_{peak}=1.5$) in our sample and we regard its spectral index as a lower limit ($\alpha_{min}$). 
 }
  \label{radio_image_hchii_2}
 \end{figure*}

Theoretically, the extremely small physical size of \hchii\ regions ($\lesssim 0.03$ pc) makes them highly likely to be point sources even in these interferometric radio surveys. However, most \hchii\ regions are located in complex star formation regions and/or surrounded by more diffuse ionized gas \citep{Sewilo2004ApJ,Sewilo2011ApJS}, which tends to result in more extended emission, particularly at lower frequencies. Thus, the 1.4 GHz flux density of \hchii\ regions may be contaminated by their environment and exhibit more extended emission than at 5 GHz. We give two examples of this in Figure~\ref{radio_image_hchii_1} and \ref{radio_image_hchii_2}, which show the morphology of two known \hchii\ regions at 1.4 GHz (white contours) and 5 GHz (lime contours). The lower frequency emission traced by MAGPIS is clearly extended in both sources --- by a neighbouring \uchii\ region in \hchii\  G45.07$+$0.13 NE (Figure \ref{radio_image_hchii_1}) and by extended emission near \hchii\ W49A G (Figure \ref{radio_image_hchii_2}).

Therefore, we consider that the spectral indices of extended sources in the sample should be strictly considered as lower limits, and we have identified them as such in our results. 
In total, we obtain 256 point sources with reliable spectral indices and 154 sources that are extended at 1.4 GHz with lower limits to their spectral index. 

\subsection{1.4 GHz dropouts in CORNISH: spectral index lower limits}
\label{sect:method}

As well as deriving the spectral index from measurements at both 1.4 and 5 GHz, it is also possible to determine a lower limit to the spectral index as described in Equation \ref{alpha_min}. In order to derive $\alpha_{\rm min}$ we need to identify sources in the CORNISH catalog that do not possess counterparts in any of the 1.4 GHz surveys (aka 1.4 GHz ``dropouts''). 
We do this by searching for sources in the CORNISH catalog that do not lie within 30\arcsec\ of \emph{any} source within the THOR catalog and 10\arcsec\ of \emph{any} source within the MAGPIS and White 2005 catalogs (the differing search radii result from the different beam sizes of the three 1.4 GHz surveys). 
That is to say, for each CORNISH source, we search for a THOR source inside a circle of 30$^{''}$ radius and a MAGPIS/White2005 source within 10\arcsec\ radius. If no 1.4 GHz sources lie inside this circle, 
the CORNISH source is preliminarily believed to be only detected at 5 GHz by CORNISH. 
An initial list of 173, 102, and 1323 dropout sources is obtained from THOR, MAGPIS, and White2005 respectively by this step. 

As a secondary constraint, to avoid associations with large, extended THOR/MAGPIS/White2005 sources 
(which may contain sources extending over several arc minutes), we also require each CORNISH source 
found in the first step to have a separation from each THOR/MAGPIS/White2005 source greater than 1.2 times 
the THOR/MAGPIS/White2005 source angular diameter. 
With this step, we remove the very large angular diameter THOR/MAGPIS/White2005 sources that may appear as ``over-resolved'' by the higher resolution CORNISH survey.
This constraint reduces our $\alpha_{\rm min}$ sample to 123, 101, and 1245 sources for THOR, MAGPIS, and White2005 respectively.  

Next, we create a unique catalog by merging the individual THOR, MAGPIS and White2005 dropouts samples and removing the duplicates caused by the overlapping survey regions. Where measurements were taken of a CORNISH source by more than one of THOR, MAGPIS or White2005, we remove the measurements with the highest noise values and keep the most sensitive measurement (generally this belongs to the THOR or MAGPIS surveys). 
Again we must also take into account the fact that only a portion of the MAGPIS images 
have a published catalogue, as described in Section \ref{sect:matches}. For each qualifying CORNISH source located between $\ell$ = 32$-$48.5\degr, we accordingly inspected 
corresponding image cutouts from MAGPIS\footref{second_footnote} to determine whether a 1.4 GHz counterpart had been detected by MAGPIS. If no 1.4 GHz counterpart had been detected, we measured the RMS noise at the position of the CORNISH source to determine an upper limit to the 1.4 GHz flux. Overall a total of 188 CORNISH sources were found to not have identifiable 1.4 GHz counterparts 
in the THOR, MAGPIS or White2005 surveys (118 THOR dropouts, 6 MAGPIS dropouts, and 64 White2005 dropouts).
  
Finally, we calculated a lower limit to the spectral index  ($\alpha_{\rm min}$) for each source using the CORNISH flux density and a 5$\sigma$ upper limit to the 1.4 GHz flux at the CORNISH position obtained from the noise maps of the THOR, MAGPIS and White2005 surveys. 
Again as we are principally interested in positive spectrum sources, we filter out those sources with $\alpha_{\rm min} < 0$ 
and list the remaining 124 positive spectrum objects in Appendix Table\,\ref{summary_positive_objects}.

\subsection{Selection of \hii\ regions from the sample}
\label{sect:selecting} 
Using the methods outlined in the previous section, we have identified a total of 256 radio sources whose spectral index $\rm \alpha-d_{\alpha} > 0$ 
and a further 278 positive spectrum objects with a minimum spectral index $\alpha_{\rm min} > 0$, obtaining a initial sample of 534 positive spectrum objects. We present this sample in Appendix Table~\ref{summary_positive_objects}. However, radio sources with positive spectra are not just limited to \hchii\ regions, and this sample is likely to contain radio galaxies and planetary nebulae as well as \hii\ regions. We must also consider that the radio surveys used here were not observed simultaneously and so radio sources whose flux has varied between the 1.4 GHz and 5 GHz measurements may result in the derivation of an incorrect spectral index. Caution must thus be applied to the interpretation of the resulting sample. 

\begin{figure*}
 \centering
 \begin{tabular}{cc}
 \includegraphics[width = 0.50\textwidth]{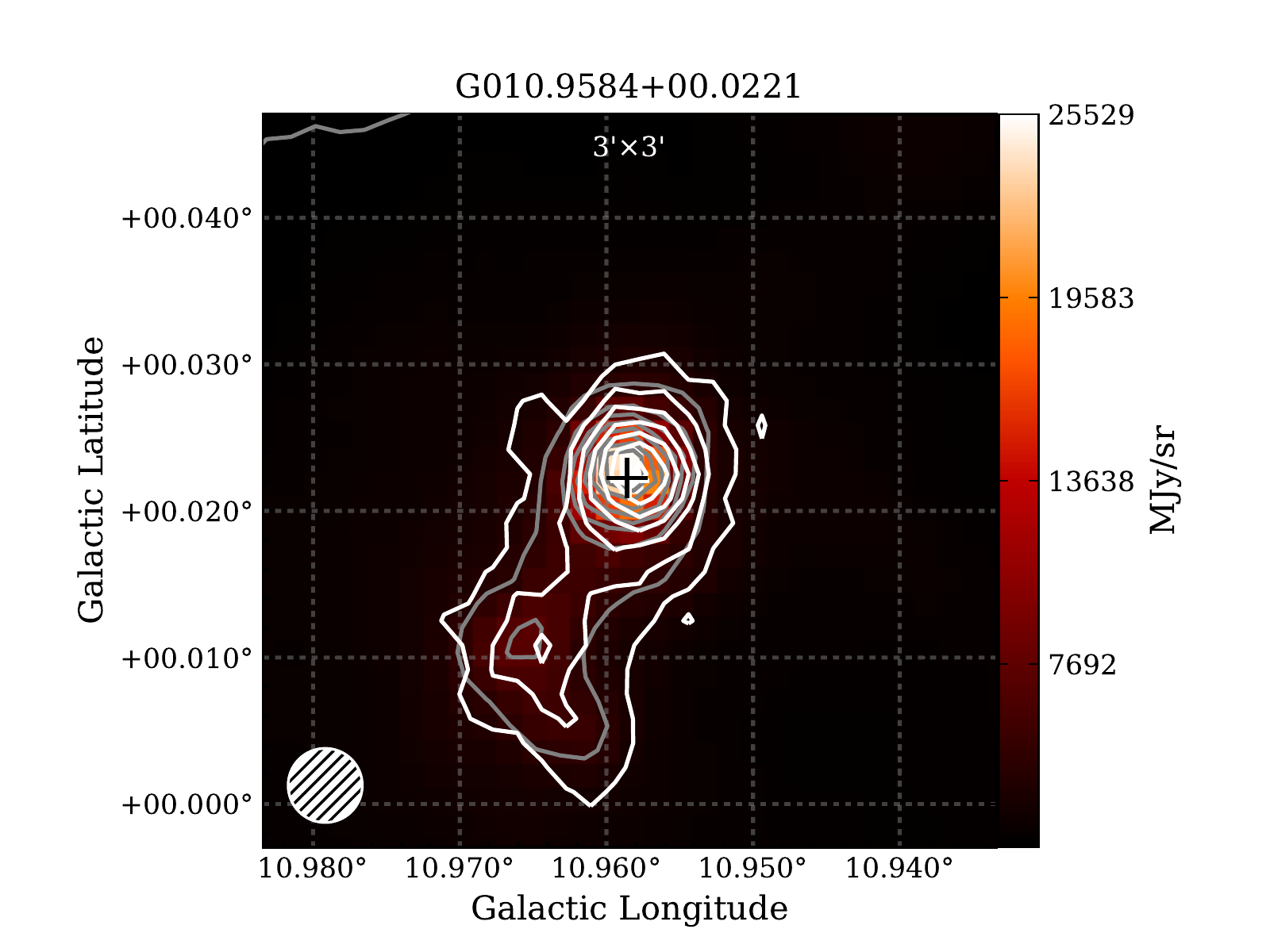}&
 \includegraphics[width = 0.50\textwidth]{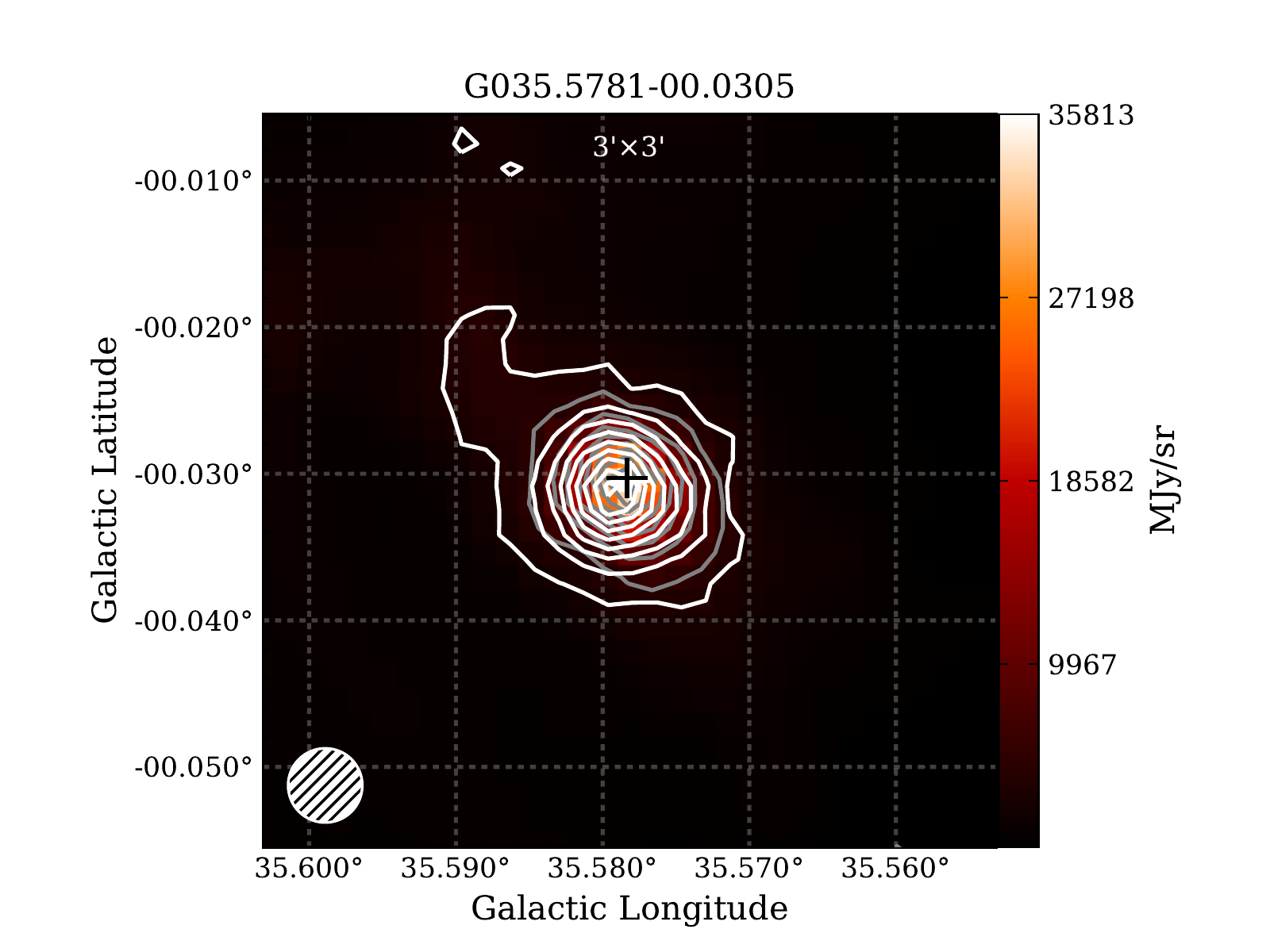}\\  
 \end{tabular}
 \caption{Left: the 250\,$\mu$m image from Hi$-$GAL overlaid with 870\,$\mu$m emission from ATLASGAL 
 for  \hchii\  region G10.96$+$0.01 W (CORNISH counterpart: G010.9584$+$00.0221). Right: the 250$\mu$m image from Hi$-$GAL overlaid with 870\,$\mu$m emission from ATLASGAL for \hchii\ region G35.58$-$0.03 (CORNISH counterpart: G035.5781$-$00.0305).  
 Grey and white contours on each image are determined by a dynamic range power-law fitting scheme \citep{Thompson2006AA} 
 and show 250$\mu$m and 870$\mu$m emission respectively.
 The beam of Hi$-$GAL 250$\mu$m (18\arcsec, similar to the ATLASGAL beam of 19\arcsec) is shown in the lower left of each image. 
 The cross indicates the position of the source at 5\,GHz from CORNISH.}
  \label{250um_870um_hchii}
\end{figure*}

In order to select \hii\ regions from our sample of positive spectrum objects, we use the Hi-GAL and ATLASGAL surveys to identify those radio sources that are also associated with far-infrared and sub-mm emission. Our approach mirrors two recent studies to identify young and ultracompact \hii\ regions \citep{Cesaroni2015AA579A,Urquhart2013MNRAS}. The appearance of each positive spectrum radio source was inspected in ATLASGAL 870 $\mu$m and Hi-GAL 250 $\mu$m cutout images, and radio sources that are positionally associated with compact 870 and 250 $\mu$m emission were identified. These radio sources are likely to be located within molecular cloud clumps, and so represent young \hii\ regions still in their embedded phase. We show two examples of known \hchii\ regions and their far-infrared and sub-mm morphologies in Figure \ref{250um_870um_hchii} to illustrate their association with compact FIR and sub-mm emission. It is possible that a small number of these sources are chance alignments of background radio galaxies with foreground ATLASGAL and/or Hi-GAL sources. However we expect the number of these alignments to be small, as the total number of chance alignments for CORNISH was estimated by \citet{Urquhart2013MNRAS} to be 14$\pm$4, and the majority of radio galaxies exhibit a negative spectral index. Hence, the number of chance alignments in our sample is expected to be statistically insignificant.

Of the 534 objects in our positive spectrum sample, 26 sources lie outside of the ATLASGAL and Hi-GAL survey regions and so are excluded from further analysis. A total of 120 objects from the remaining sample of 508 are found to be positionally associated with 870 $\mu$m and 250$\mu$m emission and are thus highly likely to be young \hii\ regions. We list this sample of 120 young positive spectrum \hii\ regions in Appendix Table\,\ref{summary_young_hii}. 
We compare our sample of young positive spectrum \hii\ regions to the young \hii\ region and \uchii\ regions samples of \citet{Cesaroni2015AA579A} and \citet{Urquhart2013MNRAS}. Unsurprisingly, due to the similar selection process for embedded sources, we find an almost one-to-one match: out of 120 \hii\ regions in our positive spectrum sample, 113 objects correspond to \citet{Cesaroni2015AA579A} young \hii\ regions or \citet{Urquhart2013MNRAS} \uchii\ regions. The only disparity between the numbers of objects in common arises because \citet{Cesaroni2015AA579A} merge individual radio sources in their study that are closer together than 11.5\arcsec\ into a single artificial source.  
In summary, by combining ATLASGAL and Hi-GAL surveys with our positive spectrum radio source sample, we have identified 120 embedded \hii\ regions with positive radio spectra. 

\subsection{Recovery of known \hchii\ regions}
\label{sect:recovery} 

As a critical test of our method, we examine the recovery of the known \hchii\ regions presented in Table \ref{summary_hchii}. Of these 16 known \hchii\ regions, 11 lie within the CORNISH survey region. With two exceptions, $\rm M17-UC1$ and $\rm W51e2^{a}$, 9 \hchii\ regions are recovered by the CORNISH survey. However, when applying our method to identify positive spectrum radio sources, we find that 5 out of these 9 are not recovered.

The reason why we do not recover these sources is that they are all located within large complex regions with extended emission at 1.4 GHz, and their 1.4 GHz flux is seriously affected by the emission from their surroundings. Three \hchii\ regions (G34.26$+$0.15B, G24.78$+$0.08A1, G28.2$-$0.04 N) display strong and extended counterparts at 1.4 GHz which dominate the 5 GHz emission, resulting in an overall negative spectrum. Two \hchii\ regions (W49AA/AB) have no separate identified 1.4 GHz counterparts. 

Our method is successful at recovering known \hchii\ regions that are not located in complex environments, and as such, complements the existing discovery space where all the known examples of \hchii\ regions have been serendipitously discovered within larger complexes of \hii\ regions. Nevertheless, we must bear in mind that our search for potential \hchii\ regions is limited within complex regions and our sample is almost certainly a lower limit to the true number of  \hchii\ regions in the Milky Way.  Below we briefly discuss the known \hchii\ regions that were recovered by CORNISH and our search process.

\paragraph{G010.9584$+$00.0221: } 
This source is the known \hchii\ region G10.96$+$0.01 W (i.e., G010.9583$+$00.0223 in Table \ref{summary_hchii}) identified by \cite{Sewilo2004ApJ}. 
 In our study, G10.96$+$0.01 W has a derived spectral index of $\alpha_{min}=1.1$  
which is consistent with $\alpha_{1.4}^{5}\sim1.2$ \citep{Sewilo2004ApJ}. 
It has an angular size of 2.2\arcsec\ at 5\,GHz with a distance of 14\,kpc \citep{Sewilo2004ApJ}, so its 5 GHz linear diameter is $\rm \sim0.15\,pc$, 
which is slightly larger than the  1.3 cm size $\rm \sim 0.121\,pc$ measured by \cite{Sewilo2011ApJS} with synthesized beam $1.4\arcsec\,\times\,0.8\arcsec$. 
The larger size and extended emission at lower frequency may result from emission from  surrounding diffuse ionized gas as proposed by \cite{Sewilo2004ApJ} and \cite{Sewilo2011ApJS}. 

\paragraph{G035.5781$-$00.0305: }
This \hchii\ region is shown in the right panel of Figure~\ref{250um_870um_hchii}. G035.5781$-$00.0305 is closely associated with compact FIR and sub-mm emission, and also shows emission 
at 1.1mm, MIR and NIR wavelengths from BGPS, GLIMPSE and UKIDSS.
This source is resolved into two extremely close sources by the JVLA at 2\,cm and 3.6\,cm \citep{Kurtz1994ApJS659K}: a Western source G35.578$-$0.030 and an Eastern source G35.578$-$0.031. 
The Western source is the known \hchii\ region G35.58$-$0.03 discussed by \cite{Zhang2014ApJ}. 
The lower resolution observations of  CORNISH and MAGPIS do not resolve these individual sources, instead revealing a single source with the catalogue identifier G035.5781$-$00.0305 surrounded by  extended continuum emission. The  
extended  emission is more than likely caused by similar surrounding diffused ionized gas, for example like G010.9584$+$00.0221 \citep[e.g.][]{Sewilo2004ApJ,Sewilo2011ApJS}. Correspondingly, the angular size of this region is larger at lower frequency (2.5\arcsec\ at 5\,GHz implying a linear diameter $\rm \sim0.1\,pc$). We derive a lower limit to the spectral index of $\alpha_{min}=1.3$. 

\paragraph{G043.1665$+$00.0106:} 

This is the \hchii\ region W49A G (G043.1666$+$00.0110), found in the
W49A complex with other two nearby \hchii\ regions W49A A and W49A B \citep{dePree1997ApJ,DePree2004ApJ,Sewilo2004ApJ}. 
All of these \hchii\ regions are detected by CORNISH, 
but only W49A G is recovered in MAGPIS and White2005, 
with a derived spectral index of $\alpha_{min}=1.4$. 
Note that this spectral index should be strictly considered as a lower limit due to a moderately resolved 1.4 GHz MAGPIS counterpart ($S_{int} /S_{peak} = 1.5$), see Figure\,\ref{radio_image_hchii_2}. 
The 1.4$-$5GHz spectral index that we derive is consistent with a previously determined value $\alpha \sim 2$ between 22\,GHz and 43\,GHz \citep{dePree1997ApJ,DePree2004ApJ}. 
We measure an angular size of 3.68\arcsec\ at 5\,GHz with a distance of 11.4\,kpc, 
which corresponds to a 5 GHz linear diameter of $\rm \sim0.2\,pc$. 
This size is larger than the size determined at 3.6 cm  ($\rm \sim 0.061\,pc$ at a resolution of $\rm 0.8\arcsec\,\times 0.8\arcsec\ $) from \cite{dePree1997ApJ}. Again, this region shows extended emission  at 5 GHz from surrounding ionized gas.

\paragraph{G045.0712$+$00.1321:} 
This source is the \hchii\ region G45.07$+$0.13 NE (G045.0712$+$00.1322) \citep{Keto2008ApJ672,Sewilo2011ApJS}, which is also found near a fainter \uchii\ region G045.0694$+$00.1323 (offset by $\sim6\arcsec\,$) detected by CORNISH. Within MAGPIS these two objects are indistinguishable from each other. We derive a spectral index of $\alpha_{min}=0.68$, which should be considered to be a lower limit due to the source blending and possible extended nature at 1.4 GHz ($S_{int} /S_{peak}= 1.74$). 
We measure a angular size of 1.89\arcsec\ at 5\,GHz with a distance of 6\,kpc, resulting in a 5 GHz linear diameter of $\rm \sim0.05\,pc$, which is slightly larger than its size at 3.6 centimeter $\rm \sim 0.032\,pc$  with a synthesized beam $\rm  1.5\arcsec\,\times\,1.4\arcsec\,$) from \cite{Sewilo2011ApJS}.  

Overall, we find that all of the recovered \hchii\ regions display a larger angular size as measured in CORNISH compared to their discovery images at higher frequency \citep[e.g.][]{Sewilo2004ApJ,Sewilo2011ApJS}. Moreover, each of the known \hchii\ regions is seen to be extended at 1.4 GHz, with a peak to integrated flux ratio greater than 1.2. This suggests that \hchii\ regions are surrounded by lower density ionized gas and  implies that searches based on size alone may not recover new \hchii\  regions. This is analogous to the well-known extended emission that is found around \uchii\ regions \citep{Kim2001ApJ549,Ellingsen2005MNRAS357}, in which snapshot and limited $uv$ coverage observations of \uchii\ regions filtered out surrounding extended emission. \citet{Kim2001ApJ549} suggested that these \hii\ regions were in fact comprised of a hierarchical structure where UC components remained largely embedded within molecular clumps but a much wider expansion of the \hii\ region had occurred along the density gradient of the clump. This results in a hierarchy of scales for the ionized gas from ultracompact to extended. We may be seeing a similar phenomenon in our \hchii\ sample.


\section{Results $\&$ Discussion}
\label{sect:result_discussion}

\subsection{Observed Properties of the Sample}

\begin{figure*}
 \centering
  \begin{tabular}{cc}
    \includegraphics[width = 0.45\textwidth]{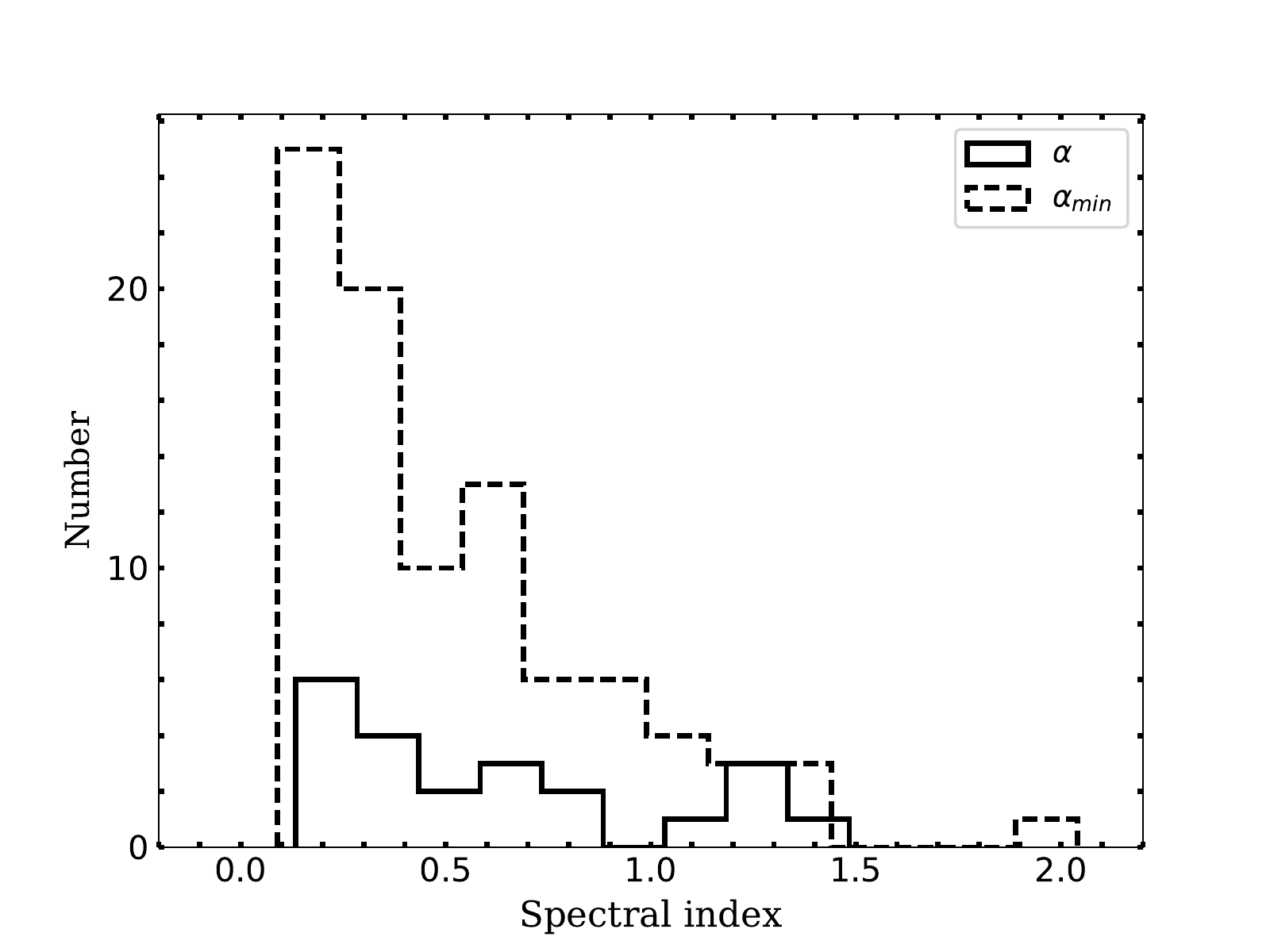}&\includegraphics[width = 0.45\textwidth]{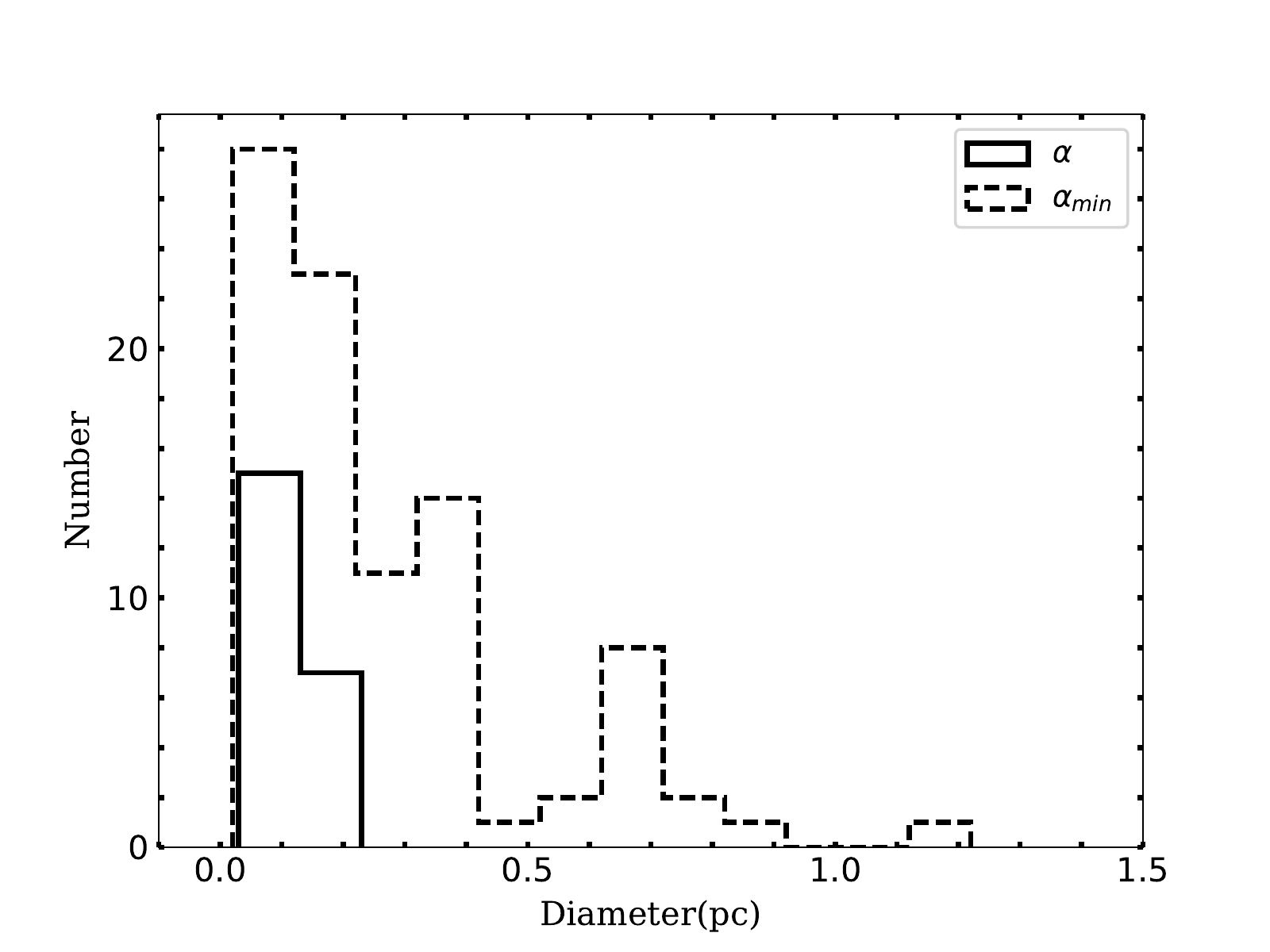}\\
  \end{tabular}
 \caption{Left: the distribution of spectral indices of \hii\ from our sample with average $\alpha = 0.6\pm0.4$ and average $\alpha_{min} = 0.5\pm0.4$. The bin size is 0.15 dex.
 Right: the distribution of  angular diameter of this sample at 5GHz with average value 
 of 0.1$\pm$0.03pc and 0.3$\pm$0.2\,pc for $\alpha$ and $\alpha_{min}$ sample respectively. The bin size is 0.1. }
 \label{fig4}
\end{figure*}

In this section, we examine the overall properties of the embedded positive spectrum \hii\ regions that we have identified. We plot histograms of their spectral indices and linear diameters in Figure \ref{fig4}. Linear diameters have been derived using the 5 GHz angular diameters measured by CORNISH \citep{Purcell2013ApJS} and the distances to the embedding molecular clumps determined by \cite{Cesaroni2015AA579A} and \cite{Urquhart2013MNRAS}.

We see that there are very few objects with purely optically thick spectral indices (i.e., $\alpha\sim2$), even taking into account those with only strict lower limits. The maximum true spectral index in our sample is $\alpha\sim$1.5, and, when lower limits to the spectral index ($\alpha_{min}$) are considered, the maximum value of $\alpha_{min}$ is $\sim$1.9. The mean values for $\alpha$ are $0.6\pm0.4$ and $\alpha_{min}$ are $0.5\pm0.4$. The majority of sources have spectral indices between 0 and 0.5. The linear diameters of the positive spectrum \hii\ regions range between 0.02--1.2 pc, with a mean of 0.1$\pm$0.03 pc for objects with true values of $\alpha$ and 0.3$\pm$0.2 pc for objects with lower limits to their spectral index.

\begin{figure*}
 \centering
\includegraphics[width = 0.7\textwidth]{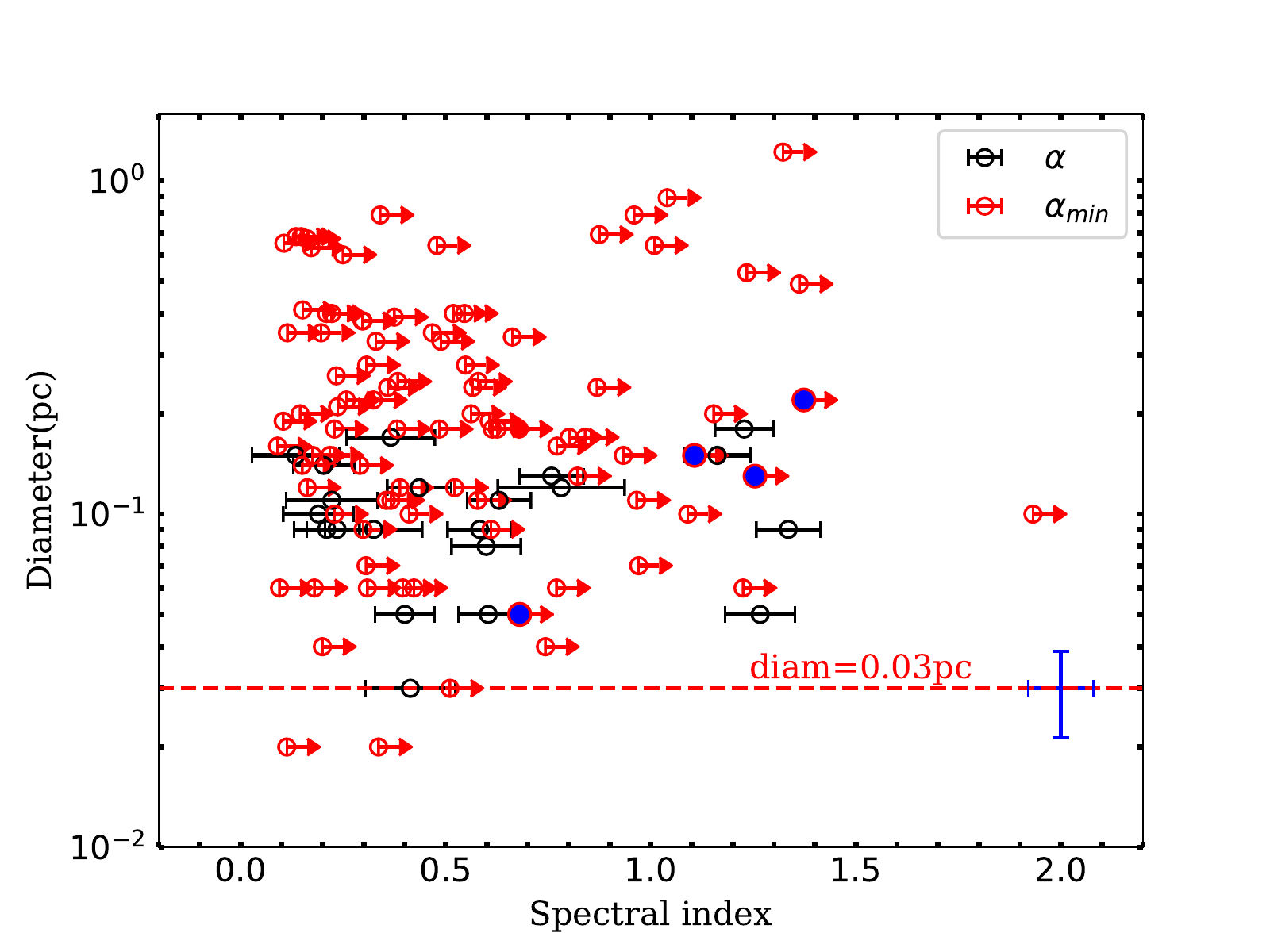}\\
 \caption{Spectral index ($\alpha$) or the lower limit of the spectral index ($\alpha_{min}$) versus the linear diameter at 5\,{\rm GHz} of our sample. 
 The filled blue circles show the four known \hchii\ regions recovered by this method. Characteristic errors for linear diameters show in blue in lower-right of the plot.  
 Rightward pointing arrows represent a lower limit of the spectral index $\alpha_{min}$. No obvious relation between spectral indices and linear diameter for $\alpha$ sample 
 as well as $\alpha_{min}$ sample, however, it is not possible to obtain real trends for $\alpha_{min}$ sample.}
 \label{diam_spc_dist}
\end{figure*}

These two distributions are combined in Figure \ref{diam_spc_dist}, where the spectral index is plotted against the linear diameter of the embedded positive spectrum \hii\ regions. The canonical diameter for \hchii\ regions is shown by a horizontal dashed line at 0.03 pc. It is immediately obvious from Figure\,\ref{diam_spc_dist} that there are \emph{no} positive spectrum \hii\ regions that fulfil the \citet{Kurtz2005IAUS} definition of diameter $\le 0.03$ pc and spectral index $\sim$ 2. Note that the exact Kurtz definition is based on emission measure, but the proposed emission measure for \hchii\ regions of $\gtrsim 10^{10}$ pc\,cm$^{-6}$ would result in $\alpha \simeq 2$ (assuming a constant density structure which is almost certainly not the case as seen in Section \ref{sect:recovery}). We do see objects with spectral indices greater than 1, but most of these have linear diameters $\geq 0.05$ pc which is more consistent with the literature definition of the ultracompact \hii\ region. The two small diameter sources ($d\leq 0.03$ pc) in our sample both have lower limits to their spectral index that are below one.

Interestingly, we also see larger $\sim$pc diameter \hii\ regions that fit the definition of compact to classical \hii\ regions, but with positive radio spectra rather than the $-0.1$ spectral index expected from optically thin emission. This suggests that these larger \hii\ regions have density gradients indicating a mix of optically thin and thick components along the line of sight.

The lack of positive spectrum \hii\ regions that fulfill the canonical picture for \hchii\ regions is puzzling. On the one hand, this may indicate that these regions are indeed rare. But on the other, we note that the known \hchii\ regions recovered in our search all show 5 GHz linear diameters larger than that expected for \hchii\ regions (the filled blue circles in Figure\,\ref{diam_spc_dist}). Thus, the fact that we do not find any such regions fitting the expected definition may merely be the result of the definition being incorrect! If \hchii\ regions are indeed likely to be surrounded by lower density ionized gas in the same hierarchical structure, then their observed linear diameter will be a complex function of observing frequency and $uv$ coverage. Thus, a number of the positive spectrum objects that we have identified may well be extremely young \hchii\ analogs with extended halos. Further high-resolution multi-frequency and multi-configuration observations are required to study the morphology and physical properties of these objects over a range of size scales, so that we may determine whether our sample does indeed contain very young \hii\ regions.

\subsection{Comparison with Herschel and ATLASGAL selected \hii\ regions}

As mentioned in Section \ref{sect:selecting} we have compared our sample of positive spectrum \hii\ regions to those presented in \cite{Cesaroni2015AA579A} and \cite{Urquhart2013MNRAS}. These samples were all selected in a similar way,  i.e. by combining CORNISH, ATLASGAL and Hi-GAL, and this facilitates their cross-comparison. The added feature of our work is that we have determined the spectral index for our sample and so can split the \cite{Cesaroni2015AA579A} and \cite{Urquhart2013MNRAS} samples by their spectral index to explore differences between populations. As both Cesaroni and Urquhart samples have well-determined physical properties (e.g. clump mass, luminosity, and Lyman continuum flux) we can examine trends in these quantities with spectral index.

We have combined the Cesaroni and Urquhart catalogues into one sample, eliminating duplicates between the two catalogs, and resulting in a final sample containing 251 young embedded \hii\ regions drawn from CORNISH. We then cross-matched this against the CORNISH sources for which we determined a spectral index (or lower limit) in Section \ref{sect:alpha}, and against the CORNISH sources associated with known \hchii\ in Table\,\ref{summary_hchii}. This allows us to split the combined Cesaroni and Urquhart sample into two subsamples, those embedded \hii\ regions with positive (or rising) spectra and those \hii\ regions with flat or negative spectra (i.e., not-rising spectra). We find 118 \hii\ regions with rising spectra and 127 \hii\ regions with not-rising spectra. The remaining 6 \hii\ regions could not have their spectral indices determined and are excluded from further analysis.

\begin{figure*}
 \centering
  \begin{tabular}{cc}
    \includegraphics[width = 0.45\textwidth]{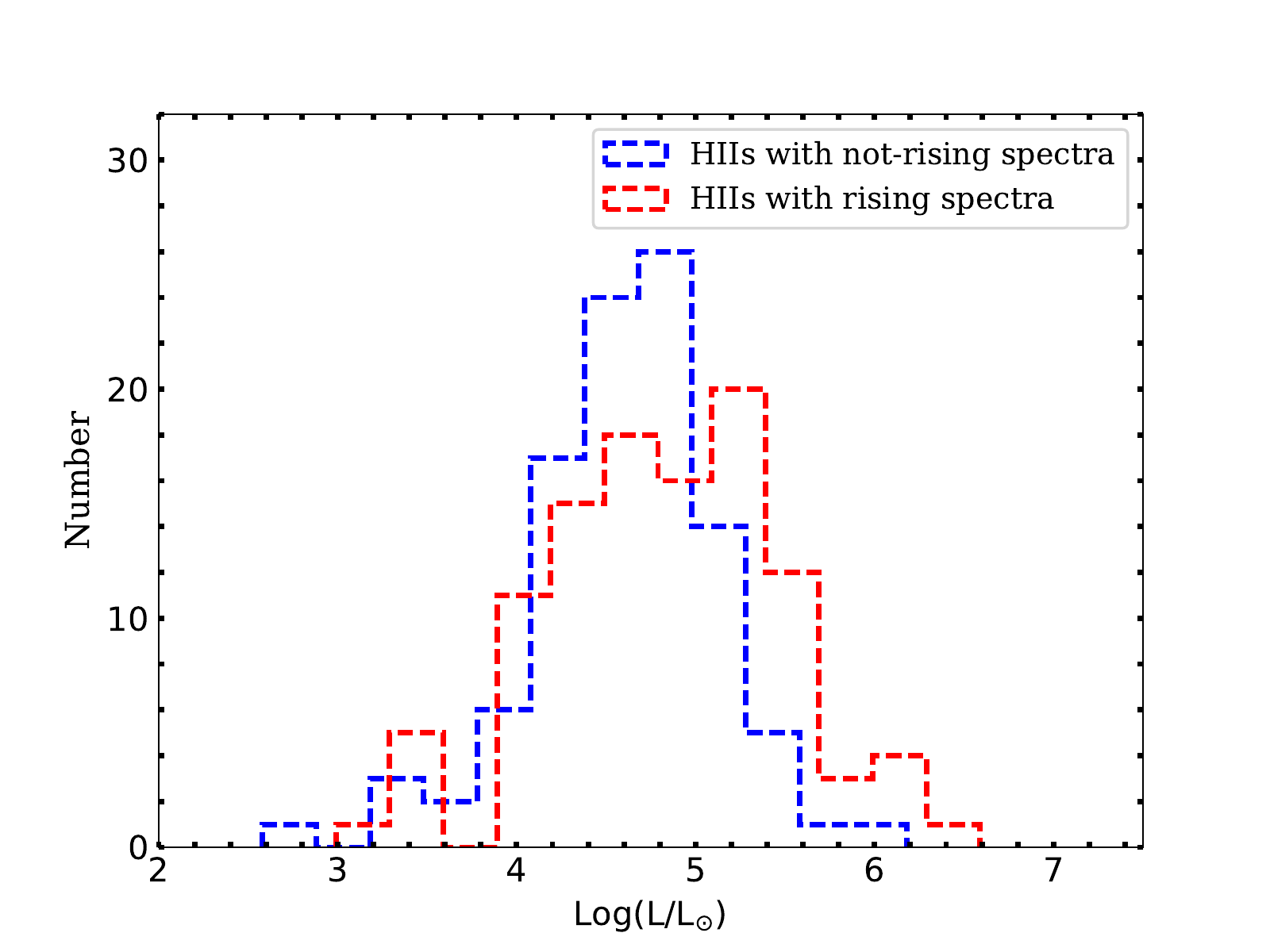}&\includegraphics[width = 0.45\textwidth]{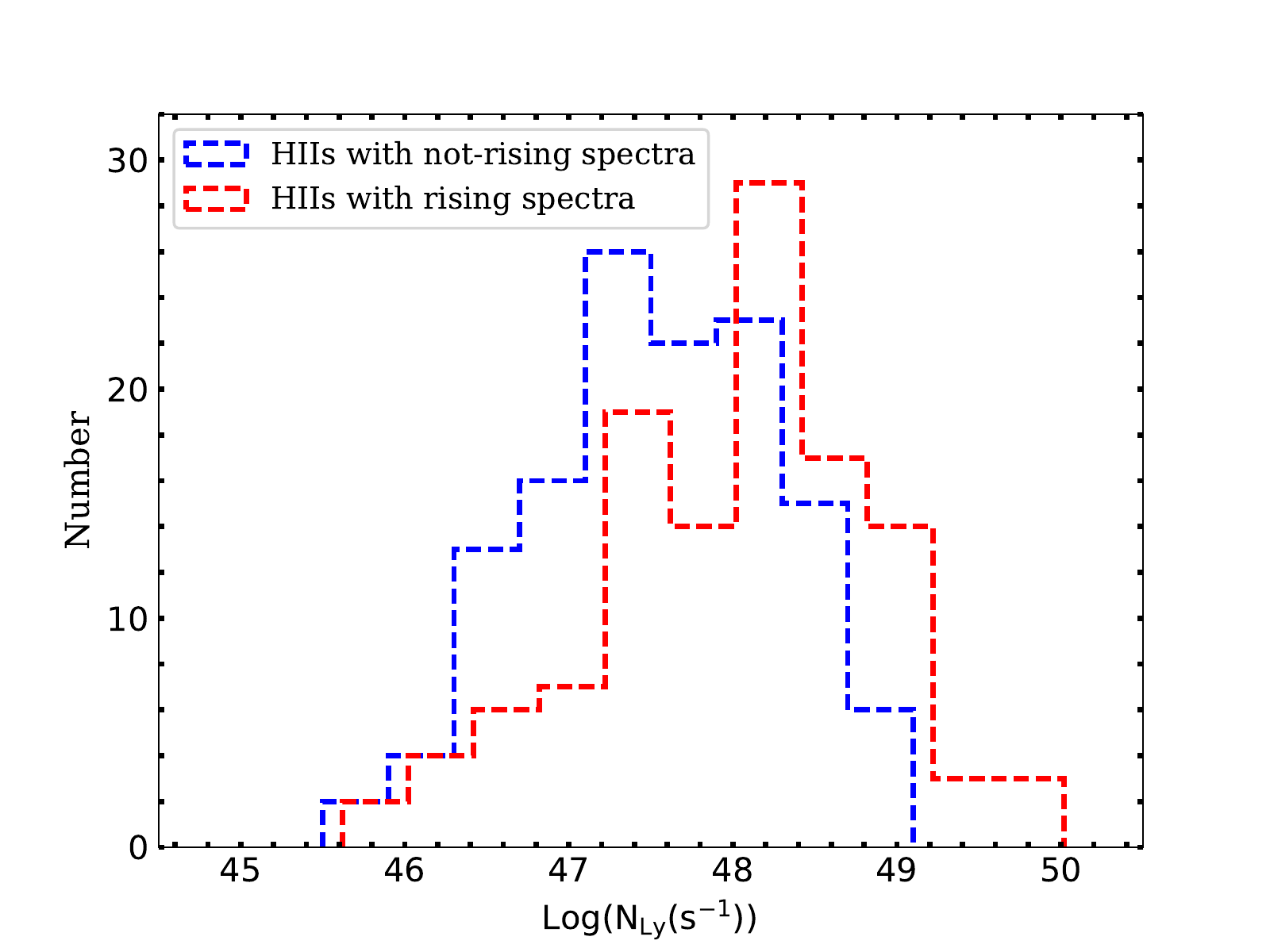}\\
  \end{tabular}
 \caption{Left-hand panel: the distribution of bolometric luminosity for rising spectra young \hii\ region and not-rising spectra young \hii\ regions, 
 with mean value of $ \rm Log(L/L_{\odot})=4.4\pm1.5$ and $\rm Log(L/L_{\odot})=3.6\pm2.0$, respectively. The bin size is 0.3.
 Right-hand panel: the distribution of Lyman continuum flux of \uchii\ regions with rising spectra and with not-rising spectra, 
 with mean value of $\rm Log(N_{Ly}(s^{-1})) = 48.0\pm0.8$ and $ \rm Log(N_{Ly}(s^{-1}))= 47.5\pm0.7$, respectively. The bin size is 0.4. }
 \label{lum_lyman_dist} 
\end{figure*}

\begin{figure*}
 \centering
  \begin{tabular}{cc}
    \includegraphics[width = 0.45\textwidth]{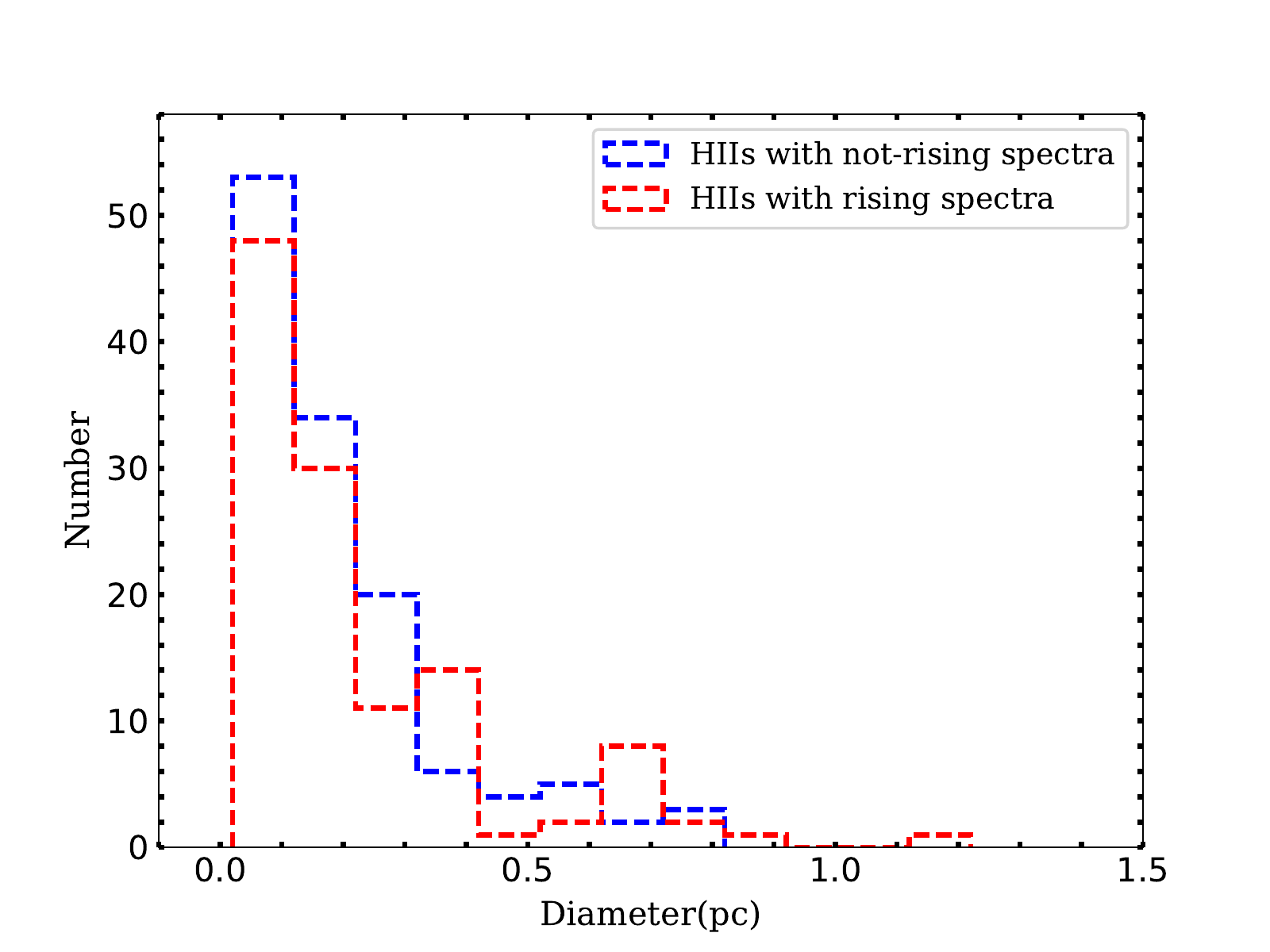}&\includegraphics[width = 0.45\textwidth]{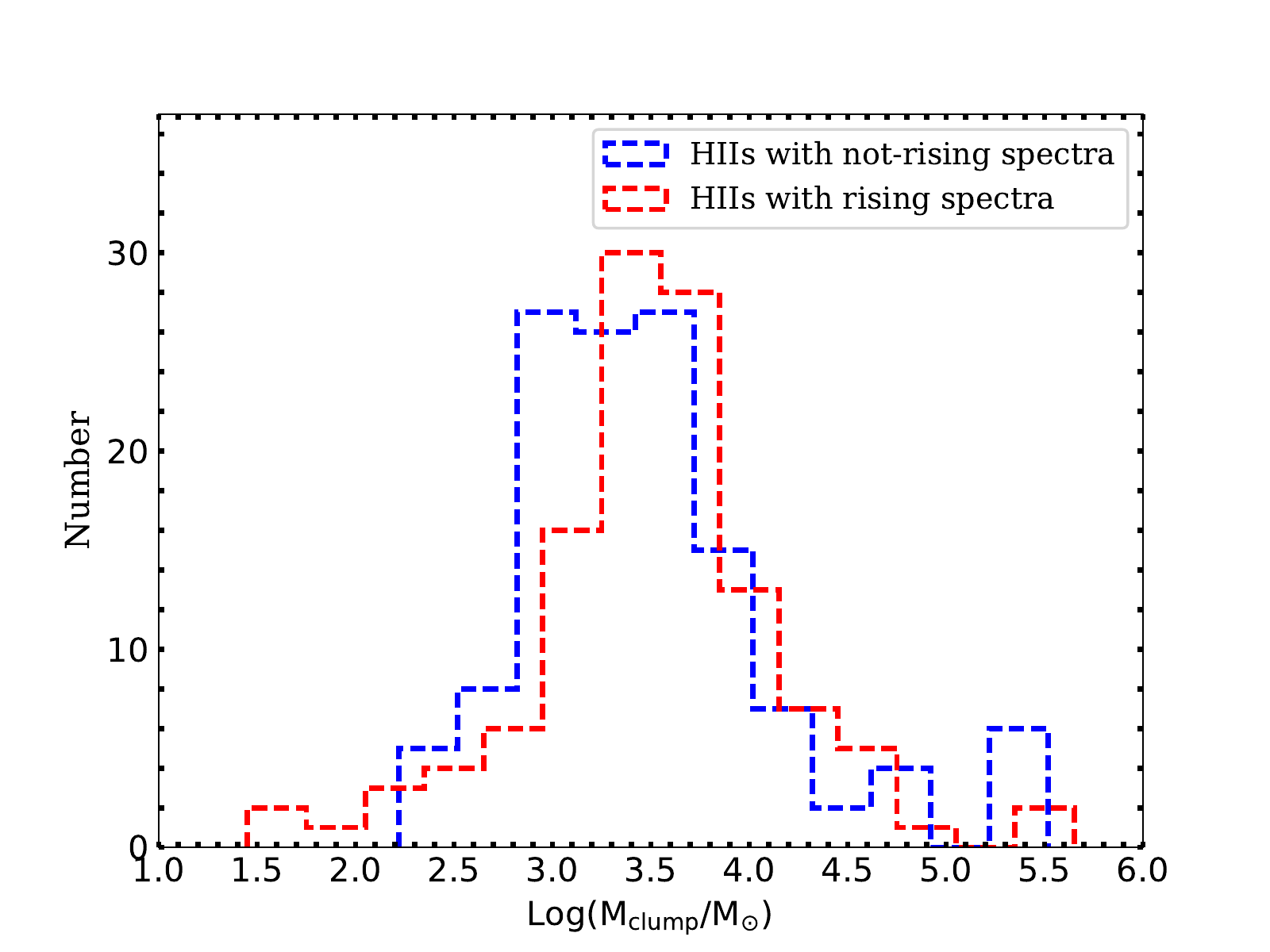}\\
  \end{tabular}
 \caption{Left-hand panel: the distribution of linear diameter for rising spectra young \hii\ region and not-rising spectra young \hii\ regions, 
 with the same mean value of $ \rm diam = 0.2\pm0.2\,pc$ for the two sample. The bin size is 0.1. 
 Right-hand panel: the distribution of clump mass of \uchii\ regions with rising spectra and with not-rising spectra, 
 with mean value of $\rm Log(M_{clump}/M_{\odot}) = 3.5\pm0.6$ and $\rm Log(M_{clump}/M_{\odot}) = 3.4\pm0.7$, respectively. The bin size is 0.3.  }
 \label{diam_mass_dist}
\end{figure*}

\begin{figure*}
 \centering
   \begin{tabular}{cc}
       \includegraphics[width = 0.45\textwidth]{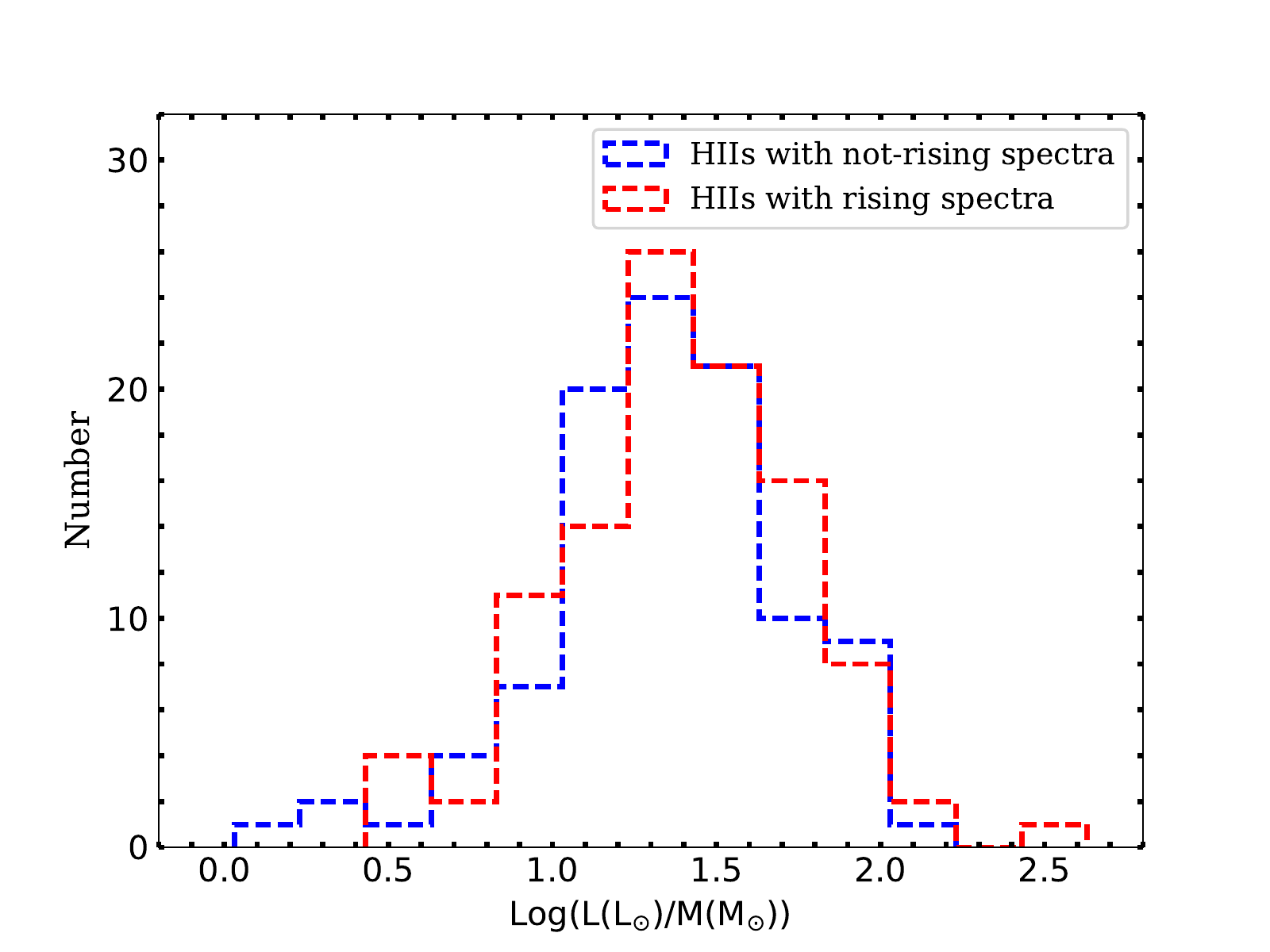} & \includegraphics[width = 0.45\textwidth]{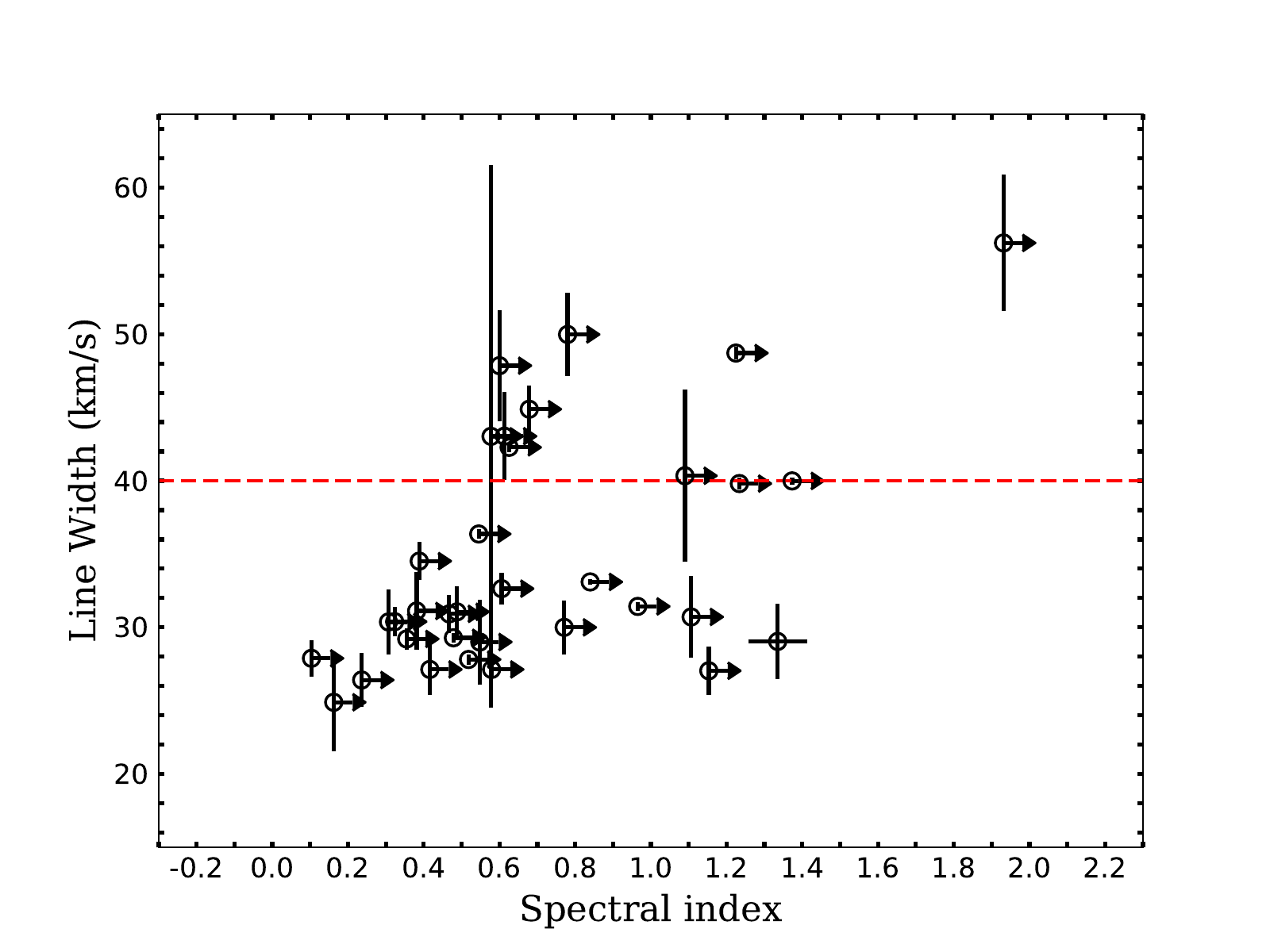} \\
    \end{tabular}
    \caption{Left-hand panel: distribution of luminosity-to-mass ratio for rising spectra young \hii\ region and not-rising spectra young \hii\ regions, 
    showing same mean value of $\rm Log(L(L_{\odot})/M(M_{\odot}))$ = $1.4\pm0.4$ for both. The bin size is 0.2. 
    Right-hand panel: spectral indices versus line widths of millimeter RRLs for 34 rising spectra \hii\ regions. Rightward pointing arrows represent a lower limit of the spectral index $\alpha_{min}$. }
 \label{lm_spc_dist}
\end{figure*}

\citet{Cesaroni2015AA579A} performed a similar although more limited analysis (see their Figure\,1) using CORNISH and MAGPIS, in order to confirm the thermal nature of the emission from their \hii\ region candidates. We find similar qualitative results to \citet{Cesaroni2015AA579A} in that roughly half of the sample show evidence for rising spectra with the remainder not-rising. The individual differences between our and the \citet{Cesaroni2015AA579A} results are due to the differing matching methods used (\citet{Cesaroni2015AA579A} use a simple 20\arcsec\  matching radius and do not consider the confusing effect of large diameter sources in MAGPIS). We examine the differences in the physical properties between rising and not-rising spectrum subsamples in Figure \ref{lum_lyman_dist} to Figure \ref{lm_spc_dist}. 

In Figure \ref{lum_lyman_dist} (left-hand panel) we show the distribution of bolometric luminosity of the two subsamples. Although the means of the two subsamples are similar with a mean $ \rm Log(L/L_{\odot})$ of $4.4\pm1.6$ for rising spectrum \hii\ regions compared to $3.6\pm2.0$ for not-rising \hii\ regions, we do see a shift towards higher luminosities for rising spectra \hii\ regions. A Kolmogorov$-$Smirnov (KS) test comparing the luminosity of the two subsamples yields a small p-value of 0.0011, and so we are able to reject the null hypothesis that the two subsamples are drawn from the same parent population.

We see a similar effect when we compare the Lyman continuum fluxes of rising and not-rising spectrum \hii\ regions, shown in Figure \ref{lum_lyman_dist} (right-hand panel). We take the values for Lyman continuum flux from \citet{Cesaroni2015AA579A} and \citet{Urquhart2013MNRAS}. Further details of the derivation can be found in these papers, but we note in passing that both studies assume optically thin emission at 5 GHz which may significantly underestimate the Lyman flux of optically thick emission. Comparing the means of the two subsample we find that the mean $\rm N_{Ly}$ for the rising spectrum \hii\ regions ($\rm Log(N_{Ly}(s^{-1})) = 48.0\pm0.8$) is moderately  larger than the not-rising spectrum \hii\ regions ($ \rm Log(N_{Ly}(s^{-1}))= 47.5\pm0.7$). A KS test of the Lyman fluxes of the two subsamples returns a p-value of $2\times10^{-5}$, and thus we are able to significantly reject the null hypothesis that the two subsamples are drawn from the same population. However, we must draw attention to the possibility of systematic bias in the rising spectrum sample due to optical depth effects. As this means that the Lyman fluxes for the rising spectrum \hii\ regions may be underestimated, the true disparity between the subsamples may be greater than we have indicated. \citet{Kim2017AA602A} compared independent radio continuum and millimeter-wave recombination line analyses for a sample of \hii\ regions, with the result that optical depth effects appeared not unduly to affect the results. However, further investigation of our rising spectrum sample is needed to confirm this hypothesis.

In Figure~\ref{diam_mass_dist} and \ref{lm_spc_dist}, we show the distributions of linear diameter, clump mass, and luminosity-to-mass ratio ($\rm L/M(L_{\odot}/M_{\odot})$) for both rising spectrum and not-rising spectrum subsamples. All of these distributions are essentially indistinguishable for rising spectrum \hii\ regions and not-rising spectrum \hii\ regions. For the linear diameter, the respective means for rising and not-rising subsamples are identical at $0.2\pm0.2$ pc and a KS test is unable to reject the null hypothesis with a p-value of 0.7. For clump mass we find respective means of $\rm \log\,M_{\rm clump} = 3.5\pm0.6\,{\rm M_{\odot}}$ and $\rm \log\,M_{\rm clump} = 3.4\pm0.7\,{\rm M_{\odot}}$ for rising and not-rising samples, and a KS test is again unable to reject the null hypothesis that the clump masses of each subsample are drawn from the same population. Finally, for the luminosity-to-mass ratio ($\rm L_{\rm bol}/M_{\rm clump}$) we find identical mean values for both subsamples ($\rm mean \rm \log\,L(L_{\odot})/M(M_{\odot}$ = $1.4\pm0.4$) and a KS test is unable to reject the null hypothesis that $\rm L_{\rm bol}/M_{\rm clump}$ values are drawn from the same population.

On balance, we find that the subsample of rising spectrum \hii\ regions tend to have higher bolometric luminosity and Lyman continuum fluxes but are not of significantly different linear diameter or found in clumps of different mass or luminosity-to-mass ratio than the not-rising spectrum \hii\ regions. This suggests that rising spectrum \hii\ regions may result from higher luminosity (and hence higher mass) stars with larger Lyman continuum fluxes. However, their similar linear diameters and luminosity-to-mass ratios imply an evolutionary status that may be much the same between the subsamples. 
The peak luminosities of $\rm \sim 10^{5}$  L$_{\odot}$ for the two subsamples of \hii\ regions (Figure \ref{lum_lyman_dist}) is consistent with the result in \citet[see their Figure 8]{Davies2011MNRAS972D} who discussed the relative number of \hii\ regions as a function of luminosity based on data from both simulation and observation. In Figure \ref{diam_spc_dist}, we see little evidence that the rising spectrum (i.e. potentially young, dense and optically thick) \hii\ regions are consistent with the canonical \hchii\ description, which is observed by \citet{Urquhart2013MNRAS}.

We also investigate the millimeter-wave recombination line properties of our sample of positive spectrum \hii\ regions by cross-matching against the aforementioned study of \citet{Kim2017AA602A}. This sample of \hii\ regions are selected from ATLASGAL clumps observed in millimeter-wave recombination lines, and we identify common objects in our two samples by matching common ATLASGAL clumps. In total, we find 35 \hii\ regions in the positive spectrum sample that are associated with millimeter-wave recombination line detections from \citet{Kim2017AA602A}, after removing one clump that is associated with two positive spectrum \hii\ regions. We plot the recombination line widths against our derived spectral indices ($\alpha$ and $\alpha_{\rm min}$) in the right-hand panel Figure \ref{lm_spc_dist}. 

Figure \ref{lm_spc_dist} shows a generally positive trend between the recombination line width and spectral index, in that \hii\ regions with broader line widths have larger spectral indices. However, it is difficult to confirm this as a genuine relationship between line width and spectral index as many of the spectral indices plotted here are strict lower limits rather than true values. We also indicate in Figure \ref{lm_spc_dist} the commonly-chosen dividing line between \hchii\ regions and \uchii\ regions at a line width of 40\,km\,s$^{-1}$. While objects with line widths in excess of 40 km\,s$^{-1}$ do display larger spectral indices, there is no clear distinction between the two.

\subsection{Implications for the frequency of \hchii\ regions and the formation of massive stars}
\label{sect:implication}

The over-riding feature of our results is that \hchii\ regions following the canonical definition of \citet{Kurtz2005IAUS} are not common in our sample. This was also noted by \citet{Kim2017AA602A} who found no hypercompact or high emission measure \hii\ regions in their 5 GHz selected sample. \citet{Kim2017AA602A} explain the lack of \hchii\  regions in their sample due to the observational bias that we discussed in Section \ref{sect:intro}. However, given that we have carried out a wide-area survey, demonstrated that CORNISH is able to recover approximately half of the known \hchii\  regions in Table \ref{summary_hchii} and that the recovered 5 GHz linear diameters of these objects are larger than the canonical definition of \hchii\  regions we do not feel that this is the most likely explanation. The true picture of \hchii\  regions is that they are likely  to be comprised of a hierarchical structure similar to \uchii\ regions \citep[e.g.][]{Kim2001ApJ549}, with highly compact dense, high emission measure ``cores'' surrounded by lower density, lower emission measure ``halos''. The distribution of spectral indices in our sample is indicative of this structure, with the majority of indices falling between 0 and 1 which implies mixed optically thin and thick emission along the line of sight. Thus the perhaps simplistic definition of \hii\ regions based on size may complicate matters and should be revised to take account of the complex structure of these objects.

Nevertheless, in Figure \ref{lum_lyman_dist}, the rising spectrum \hii\ regions (i.e., potentially young and dense) are more likely to be of higher luminosity and have higher Lyman continuum fluxes than the not-rising spectrum \hii\ regions. \citet{Urquhart2013MNRAS} also found that the most highly luminous \hii\ regions are amongst the largest. \hii\ regions show an excess of Lyman continuum that the measured values are larger than the theoretical prediction, which cannot be easily explained \citep{Lumsden2013ApJS208,Urquhart2013MNRAS}. Further research is needed to understand these differences of luminosity and Lyman continuum flux between the two subsamples \hii\ regions. 

It is difficult to assess how complete our observations are of the potential hypercompact population of \hii\ regions. We find that CORNISH is able to recover roughly half of the known sample of \hchii\ regions and so one might naively assume that there are perhaps a factor 2 more \hchii\ regions to be discovered in the CORNISH survey region. However, it is clear that \hii\ regions with positive spectra are in fact common, as roughly half of the \citet{Cesaroni2015AA579A} and \citet{Urquhart2013MNRAS} catalogs of compact and ultracompact \hii\ regions have positive spectral indices. This may have implications as the ionized gas properties that have been derived in many studies of \hii\ regions to date have assumed that the continuum emission is optically thin, rather than the mixture of optically thick and thin components that our distribution of spectral indices implies. Detailed multi-frequency and multi-configuration observations that can reveal the ionized structure of these regions on a range of scales are required to further investigate their nature and examine their relationship to the early phases of massive star formation.

\section{Summary and conclusions}

We have carried out the largest and most unbiased search for hypercompact \hii\ regions to date by combining radio surveys at 1.4 and 5 GHz (THOR, CORNISH, MAGPIS and White2005) with far-infrared and sub-mm Galactic Plane surveys (Hi-GAL and ATLASGAL). We obtain a sample of 534 objects with a 1.4 to 5 GHz spectral index greater than zero, listed in Appendix Table \ref{summary_positive_objects}. 256 of these objects were detected at 5 GHz and as point sources at 1.4 GHz which means that we could determine true values of their spectral index, whereas the remaining 278 objects  have upper limits at 1.4 GHz or were found to be moderately extended at 1.4 GHz and thus have strict lower limits to their spectral index. We identified \hii\ regions in this sample using ATLASGAL and Hi-GAL surveys in a similar manner to the recent studies of young and ultracompact \hii\ regions by \citet{Cesaroni2015AA579A} and \citet{Urquhart2013MNRAS}. We found a total of 120 \hii\ regions with positive radio spectral indices, shown in Appendix Table \ref{summary_young_hii}. Among the 120 positive spectra \hii\ regions, 35 have archival Radio Recombination Line (RRL) observations, see the right-hand panel of Figure\,\ref{lm_spc_dist}. Twelve out of the 35 \hii\ regions show broad RRL line-width $\rm \Delta V \gtrsim 40\,km\,s^{-1}$, listed in Table\,\ref{promising_hchii}. Four of the 12 sources are known HCHII regions in Table\,\ref{summary_hchii}, and follow-up JVLA observations for the rest \hii\ regions have been carried out to determine their nature. The physical properties of the 120 rising spectra \hii\ regions were examined and compared to the \citet{Cesaroni2015AA579A}, \citet{Urquhart2013MNRAS} and \citet{Kim2017AA602A} samples of \hii\ regions. We draw the following conclusions:

\begin{enumerate}

\item We find \emph{no} objects in our resulting positive spectrum \hii\ region sample that match the canonical definition of \hchii\ regions given in \citet{Kurtz2005IAUS}, i.e. with linear diameter $\le 0.03$ pc and with a spectral index $\simeq 2$. The majority of our positive spectrum \hii\ regions have diameters more than 0.03 pc and with spectral indices between 0.5--1.0. 

\item We recover roughly half of the known \hchii\ regions in the CORNISH survey. However, these objects are generally resolved at 5 GHz with larger diameters than seen in their higher frequency discovery observations. Combined with spectral indices that indicate mixed optically thick and thin components along the line of sight, we suggest that \hchii\ regions have a hierarchical structure analogous to \uchii\ regions. Multi-frequency, multi-resolution radio observations are required to confirm this hypothesis. The canonical definition of \hii\ regions based on linear diameter may perhaps need to be updated to reflect its structure. 

\item We see a general trend between spectral index and the line width of mm-wave recombination lines observed by \citet{Kim2017AA602A}, in that objects with higher spectral indices tend to show higher line widths. However, this trend is still inconclusive as many of the spectral indices for the source in \citet{Kim2017AA602A} sample are lower limits. Further higher frequency radio continuum observations are required to confirm this trend.

\item In a combined sample drawn from \citet{Cesaroni2015AA579A} and \citet{Urquhart2013MNRAS} we find that roughly half of these \hii\ regions have positive spectral indices. \hii\ regions with a positive (i.e. rising) spectrum are found to be statistically more luminous and with higher Lyman fluxes than \hii\ regions with negative or flat (i.e., not-rising) spectral indices. This suggests that rising spectrum \hii\ regions are associated with more luminous and massive stars. We find no evidence for differences in the linear diameter of rising and not-rising spectrum \hii\ regions, nor in the mass of their embedding clumps or their luminosity-to-mass ratios.

\end{enumerate}
  
  \begin{table*}
 \setlength{\tabcolsep}{5pt}
\scriptsize
 \caption {\rm 12 Rising Spectra \hii\ regions with Broad Radio Recombination Lines ($\rm FWHM \gtrsim 40\,km\,s^{-1}$) in Our Sample}
  \begin{tabular}{lllcccllll}
  \hline
   \hline
   Gname &	Flux$_{5\,GHz}$	& Ang.\,(5\,GHz) &Flux$_{1.4\,GHz}$    & Spectral Index  &  Associated clump &$\rm \Delta V(RRL)$	& Dist. &  Diam.(5\,GHz)\\
CORNISH & (mJy) &	($\arcsec$) &	(mJy) &	($\alpha_{1.4}^{5}$) & (AGAL) & ($\rm km\,s^{-1}$) & (kpc)  &  (pc)\\
\hline
  G010.9584+00.0221$\dagger$ &   195.97$\pm$18.33   &     2.20\arcsec  &     47.92$\pm$0.40  &     1.11$^{\star}$ &    AGAL010.957+00.022 &    43.8(H92$\alpha$) &    14     &    0.15   \\ 
  G030.5887-00.0428 &   92.37$\pm$8.33     &     1.79\arcsec  &     7.90$\pm$0.48  &     1.93$^{\star}$  &    AGAL030.588-00.042 &    56.2(H40$\alpha$) &    11.8     &    0.1    \\ 
  G030.7197-00.0829 &   969.33$\pm$96.01   &     4.59\arcsec  &     464.58$\pm$2.28  &     0.58$^{\star}$  &    AGAL030.718-00.082 &    43.0(H40$\alpha$) &    4.9      &    0.11   \\ 
  G030.8662+00.1143 &   325.47$\pm$32.96   &     3.09\arcsec  &     137.17$\pm$0.60  &     0.68$^{\star}$  &    AGAL030.866+00.114 &    44.9(H39$\alpha$) &    11.9     &    0.18   \\ 
  G033.1328-00.0923 &   378.59$\pm$34.75   &     4.02\arcsec  &     173.43$\pm$1.43  &     0.61$^{\star}$  &    AGAL033.133-00.092 &    43.0(H39$\alpha$) &    9.4      &    0.18   \\ 
  G034.2572+00.1535 &   1762.63$\pm$163.28  &     5.75\arcsec  &     370.78$\pm$3.39  &     1.22$^{\star}$  &    AGAL034.258+00.154 &    48.7(H42$\alpha$) &    2.1      &    0.06   \\ 
  G035.5781-00.0305$\dagger$ &   187.75$\pm$18.44   &     2.53\arcsec  &     38.05$\pm$0.97  &     1.25$^{\star}$  &    AGAL035.579-00.031 &    50.0(H42$\alpha$) &    10.2     &    0.13   \\ 
  G043.1665+00.0106$\dagger$ &   1365.68$\pm$125.16  &     3.68\arcsec  &     237.81$\pm$8.09  &     1.37$^{\star}$  &    AGAL043.166+00.011 &    40.0(H39$\alpha$) &    11.4     &    0.2    \\ 
  G045.0712+00.1321$\dagger$ &   146.67$\pm$14.65     &     1.89\arcsec  &     61.6$\pm$0.6       &     0.68$^{\star}$  &    AGAL045.071+00.132 &    40.0(H76$\alpha$) &    6.0      &    0.05   \\ 
  G045.1223+00.1321 &   2984.27$\pm$274.33  &     7.46\arcsec  &     1345.96$\pm$1.67  &     0.63$^{\star}$  &    AGAL045.121+00.131 &    42.3(H41$\alpha$) &    4.4      &    0.16   \\ 
  G032.7441-00.0755  & 7.93$\pm$1.14 & 1.78\arcsec & 0.34 & $1.09^{\star}$ & AGAL032.744-00.076  & 40.3(Hn$\alpha$)&11.7 & 0.10\\ 
  G045.4656+00.0452      &    62.26$\pm$5.79    &    1.70\arcsec &    28.88$\pm$1.32  &    0.60$\pm$0.05 & AGAL045.466+00.046    & 47.9(H39$\alpha$)& 6.0 & 0.05\\
 \hline
\hline
\end{tabular}
\begin{tablenotes}
 \scriptsize
\item These columns contain the Galactic name in CORNISH, flux density and angular scale (Ang.) of each source at 5\,{\rm GHz}, flux density at 1.4\,GHz, and the spectral index and its error, and its ATLASGAL counterparts with Galactic name, line width of RRLs, heliocentric distance (Dist.) in kpc, and linear diameter (Diam.) at 5\,GHz. Symbol $\dagger$ means the known \hchii\ regions recovered by this method. RRLs of $\rm Hn\alpha \,(n=39,40,41,42)$ from \cite{Kim2017AA602A}. References of H76$\alpha$ for \hchii\ G045.0712$+$00.1321 and H92$\alpha$ for \hchii\ G010.9584$+$00.0221 are same as Table\,\ref{summary_hchii}. Symbol $\star$ means that sources have a lower limit for their spectral index. The flux density listed for  G032.7441-00.0755 is the 1.4\,GHz RMS noise level at   the source position.
\end{tablenotes}
\label{promising_hchii}
\end{table*}


\section*{Acknowledgements}
We would like to thank the referee for their helpful and constructive comments. We acknowledge support from the NSFC~(11603039, 11473038).  
H.B. acknowledges support from the European Research Council under the Horizon 2020 Framework Program via the ERC Consolidator Grant CSF-648505. 
M.A.T.~acknowledges support from the UK Science $\&$ Technology Facilities Council via grant ST/M001008/1. 
A.Y.~would like to thank the UK Science $\&$ Technology Facilities Council (STFC) and the China Scholarship Council (CSC) for grant funding through the China-UK SKA co-training programme. 

\appendix
 \section{tables}
 \label{appendix:table}

\begin{table*}
 \centering
\scriptsize
 \caption {\rm Information of 120 young positive spectrum \hii\ regions }
  \begin{tabular}{p{3cm}p{2cm}p{2cm}p{2cm}p{2cm}p{2cm}p{2cm}}
  \hline
Name	 &	$\ell$	&	$b$	&	Flux$_{5\,GHz}$	&	Angular diameter       &Flux$_{1.4\,GHz}$    &Spectral Index   \\
Gal	       &	($\degr$)	&	($\degr$)	&	(mJy) 	&	(\arcsec)	     &	(mJy)	                          &		      \\
\hline       
 \hline
  G010.3009$-$00.1477$\dagger$  &   10.30088   &    -0.1477    &    631.39$\pm$59.30    &    5.45$\pm$0.01 &    426.18$\pm$1.98 &    0.31 \\
  G010.4724$+$00.0275$\dagger$  &   10.47236   &    0.0275     &    38.43$\pm$4.38      &    2.24$\pm$0.02 &    31.26$\pm$0.35  &    0.16 \\
  G010.6223$-$00.3788$\dagger$  &   10.62231   &    -0.37877   &    483.33$\pm$49.87    &    5.76$\pm$0.01 &    327.63$\pm$1.89 &    0.31 \\
  G010.6234$-$00.3837$\dagger$  &   10.6234    &    -0.38369   &    1952.22$\pm$176.18  &    4.64$\pm$0.00 &    571.28$\pm$1.88 &    0.97 \\
  G010.9584$+$00.0221$\dagger$  &   10.95839   &    0.02206    &    195.97$\pm$18.33    &    2.20$\pm$0.00 &    47.92$\pm$0.40  &    1.11 \\
  G011.0328$+$00.0274$\dagger$  &   11.03283   &    0.02738    &    5.69$\pm$1.06       &    1.89$\pm$0.15 &    3.71$\pm$0.28   &    0.34 \\
  G011.1104$-$00.3985$\dagger$  &   11.11043   &    -0.39851   &    305.37$\pm$28.55    &    8.36$\pm$0.01 &    253.15$\pm$0.42 &    0.15 \\
  G011.1712$-$00.0662$\dagger$  &   11.17121   &    -0.06621   &    102.17$\pm$12.73    &    10.75$\pm$0.04 &    83.15$\pm$0.32  &    0.16 \\
  G011.9368$-$00.6158$\dagger$  &   11.93677   &    -0.61577   &    1155.90$\pm$105.38  &    5.89$\pm$0.00 &    735.63$\pm$0.43 &    0.36 \\
  G011.9446$-$00.0369$\dagger$  &   11.94458   &    -0.03686   &    943.58$\pm$98.50    &    14.64$\pm$0.01 &    251.14$\pm$1.91 &    1.04 \\
  G012.1988$-$00.0345$\dagger$  &   12.1988    &    -0.03447   &    62.71$\pm$5.92      &    2.68$\pm$0.06 &    47.56$\pm$0.31  &    0.22 \\
  G012.2081$-$00.1019$\dagger$  &   12.20806   &    -0.10189   &    207.87$\pm$19.73    &    2.84$\pm$0.01 &    127.94$\pm$1.17 &    0.38 \\
  G012.4294$-$00.0479$\dagger$  &   12.4294    &    -0.04791   &    45.17$\pm$4.35      &    2.72$\pm$0.07 &    24.39$\pm$1.26  &    0.48 \\
  G012.8050$-$00.2007$\dagger$  &   12.805     &    -0.20067   &    12616.40$\pm$1120.83 &    16.23$\pm$0.01 &    4332.25$\pm$4.27 &    0.84 \\
  G012.8131$-$00.1976$\dagger$  &   12.8131    &    -0.19759   &    1500.39$\pm$147.30  &    5.43$\pm$0.01 &    907.68$\pm$4.18 &    0.39 \\
  G012.9995$-$00.3583$\dagger$  &   12.99951   &    -0.3583    &    20.14$\pm$3.70      &    3.09$\pm$0.32 &    10.52$\pm$0.28  &    0.51 \\
  G013.2099$-$00.1428$\dagger$  &   13.20989   &    -0.14281   &    946.76$\pm$87.46    &    8.35$\pm$0.01 &    437.88$\pm$3.66 &    0.61 \\
  G013.3850$+$00.0684$\dagger$  &   13.38496   &    0.06835    &    603.94$\pm$60.83    &    19.18$\pm$0.01 &    139.16$\pm$1.16 &    1.15 \\
  G014.7785$-$00.3328    &   14.77849   &    -0.33278   &    18.25$\pm$2.47      &    2.31$\pm$0.03 &    15.39$\pm$1.04  &    0.13$\pm$0.11 \\
  G016.1448$+$00.0088$\dagger$  &   16.14482   &    0.00876    &    14.76$\pm$1.55      &    1.60$\pm$0.04 &    8.75$\pm$0.19   &    0.41 \\
  G016.3913$-$00.1383$\dagger$  &   16.39128   &    -0.13827   &    124.27$\pm$15.43    &    11.74$\pm$0.04 &    40.83$\pm$0.31  &    0.87 \\
  G016.9445$-$00.0738$\dagger$  &   16.94449   &    -0.07379   &    519.34$\pm$47.78    &    3.46$\pm$0.00 &    258.51$\pm$0.34 &    0.55 \\
  G017.0299$-$00.0696    &   17.02987   &    -0.06955   &    5.38$\pm$1.06       &    2.39$\pm$0.24 &    1.99$\pm$0.37   &    0.78$\pm$0.15 \\
  G017.1141$-$00.1124$\dagger$  &   17.11407   &    -0.11236   &    17.21$\pm$2.19      &    2.83$\pm$0.16 &    14.22$\pm$0.25  &    0.15 \\
  G018.1460$-$00.2839$\dagger$  &   18.14602   &    -0.2839    &    856.18$\pm$82.85    &    23.42$\pm$0.02 &    151.22$\pm$1.93 &    1.36 \\
  G018.3024$-$00.3910$\dagger$  &   18.30241   &    -0.39103   &    1277.88$\pm$114.83  &    14.63$\pm$0.00 &    846.82$\pm$0.32 &    0.32 \\
  G018.4433$-$00.0056$\dagger$  &   18.44328   &    -0.00558   &    81.31$\pm$7.30      &    2.41$\pm$0.04 &    56.18$\pm$0.37  &    0.29 \\
  G018.4614$-$00.0038$\dagger$  &   18.46141   &    -0.00378   &    342.12$\pm$31.50    &    2.76$\pm$0.00 &    128.20$\pm$0.47 &    0.77 \\
  G018.6654$+$00.0294    &   18.66539   &    0.02935    &    5.65$\pm$0.85       &    1.74$\pm$0.10 &    3.74$\pm$0.21   &    0.32$\pm$0.12 \\
  G018.7106$+$00.0002    &   18.71061   &    0.00022     &    107.46$\pm$10.62    &    2.04$\pm$0.01 &    40.95$\pm$0.31  &    0.76$\pm$0.08 \\
  G018.7612$+$00.2630$\dagger$  &   18.76118   &    0.26298    &    51.38$\pm$4.67      &    1.79$\pm$0.03 &    26.45$\pm$0.23  &    0.52 \\
  G018.8250$-$00.4675$\dagger$  &   18.82499   &    -0.46749   &    11.41$\pm$2.17      &    2.53$\pm$0.25 &    9.08$\pm$0.35   &    0.18 \\
  G018.8338$-$00.3002$\dagger$  &   18.83384   &    -0.30024   &    131.38$\pm$13.35    &    6.72$\pm$0.01 &    108.41$\pm$0.24 &    0.15 \\
  G019.0754$-$00.2874$\dagger$  &   19.07543   &    -0.28737   &    380.69$\pm$37.06    &    8.49$\pm$0.01 &    333.71$\pm$1.81 &    0.10 \\
  G019.4912$+$00.1352$\dagger$  &   19.49123   &    0.13518    &    415.07$\pm$38.70    &    11.92$\pm$0.01 &    269.27$\pm$1.40 &    0.34 \\
  G019.6087$-$00.2351$\dagger$  &   19.60873   &    -0.23507   &    2900.88$\pm$260.93  &    13.11$\pm$0.00 &    855.57$\pm$1.90 &    0.96 \\
  G019.6090$-$00.2313$\dagger$  &   19.60899   &    -0.23126   &    259.95$\pm$26.87    &    3.84$\pm$0.01 &    126.53$\pm$1.91 &    0.57 \\
  G019.7407$+$00.2821$\dagger$  &   19.74069   &    0.28206    &    239.01$\pm$22.33    &    17.95$\pm$0.02 &    44.41$\pm$2.80  &    1.32 \\
  G019.7549$-$00.1282$\dagger$  &   19.7549    &    -0.12817   &    36.52$\pm$3.29      &    1.64$\pm$0.03 &    10.62$\pm$0.14  &    0.97 \\
  G020.0720$-$00.1421$\dagger$  &   20.07196   &    -0.14208   &    210.13$\pm$21.54    &    5.39$\pm$0.01 &    138.10$\pm$1.14 &    0.33 \\
  G020.0789$-$00.1383$\dagger$  &   20.07889   &    -0.13826   &    295.86$\pm$28.49    &    19.66$\pm$0.03 &    72.47$\pm$1.14  &    1.11 \\
  G020.0809$-$00.1362    &   20.0809    &    -0.13617   &    498.19$\pm$45.06    &    2.98$\pm$0.00 &    104.42$\pm$1.15 &    1.23$\pm$0.07 \\
  G020.3633$-$00.0136$\dagger$  &   20.36326   &    -0.01355   &    55.11$\pm$5.93      &    2.93$\pm$0.01 &    32.20$\pm$1.81  &    0.42 \\
  G021.3571$-$00.1766    &   21.35708   &    -0.17658   &    24.93$\pm$2.34      &    1.91$\pm$0.04 &    18.49$\pm$0.16  &    0.23$\pm$0.07 \\
  G021.3855$-$00.2541    &   21.38554   &    -0.25408   &    113.91$\pm$11.24    &    2.22$\pm$0.01 &    51.09$\pm$0.18  &    0.63$\pm$0.08 \\
  G021.4257$-$00.5417$\dagger$  &   21.42574   &    -0.54167   &    94.85$\pm$13.38     &    10.22$\pm$0.04 &    78.89$\pm$0.24  &    0.14 \\
  G023.2654$+$00.0765$\dagger$  &   23.26542   &    0.07647    &    88.57$\pm$9.87      &    4.58$\pm$0.02 &    55.58$\pm$2.45  &    0.37 \\
  G023.4181$-$00.3940    &   23.4181    &    -0.39403   &    26.46$\pm$2.47      &    1.77$\pm$0.03 &    18.96$\pm$0.21  &    0.26$\pm$0.07 \\
  G023.4553$-$00.2010    &   23.45526   &    -0.20096   &    14.39$\pm$1.56      &    1.71$\pm$0.05 &    2.87$\pm$0.47   &    1.27$\pm$0.09 \\
  G024.5065$-$00.2224$\dagger$  &   24.50653   &    -0.22238   &    205.57$\pm$19.72    &    6.19$\pm$0.01 &    153.69$\pm$2.77 &    0.23 \\
  G024.9237$+$00.0777$\dagger$  &   24.92371   &    0.07769    &    172.48$\pm$20.21    &    14.93$\pm$0.03 &    57.08$\pm$1.87  &    0.87 \\
  G025.3948$+$00.0332$\dagger$  &   25.39478   &    0.03324    &    296.86$\pm$27.46    &    4.64$\pm$0.01 &    203.65$\pm$2.64 &    0.30 \\
  G025.3970$+$00.5614    &   25.39695   &    0.5614     &    121.17$\pm$11.53    &    2.04$\pm$0.01 &    93.65$\pm$0.68  &    0.20$\pm$0.07 \\
  G025.3981$-$00.1411$\dagger$  &   25.39808   &    -0.14107   &    2132.24$\pm$194.31  &    8.60$\pm$0.01 &    1351.54$\pm$3.30 &    0.36 \\
  G025.7157$+$00.0487    &   25.71567   &    0.04868    &    20.79$\pm$2.96      &    2.36$\pm$0.03 &    15.67$\pm$1.00  &    0.22$\pm$0.11 \\
  G025.8011$-$00.1568    &   25.80114   &    -0.15685   &    31.95$\pm$2.96      &    1.79$\pm$0.03 &    19.20$\pm$0.20  &    0.40$\pm$0.07 \\
  G026.5444$+$00.4169$\dagger$  &   26.54436   &    0.4169     &    413.36$\pm$37.39    &    12.59$\pm$0.01 &    301.02$\pm$1.00 &    0.25 \\
  G027.2800$+$00.1447$\dagger$  &   27.27996   &    0.14468    &    428.04$\pm$42.07    &    5.75$\pm$0.01 &    370.37$\pm$0.25 &    0.11 \\
  G027.3644$-$00.1657$\dagger$  &   27.3644    &    -0.16574   &    60.14$\pm$6.13      &    2.26$\pm$0.01 &    44.95$\pm$0.21  &    0.23 \\
  G027.9782$+$00.0789$\dagger$  &   27.97822   &    0.07893    &    124.00$\pm$14.38    &    9.47$\pm$0.03 &    89.34$\pm$1.76  &    0.26 \\
  G028.2879$-$00.3641$\dagger$  &   28.28789   &    -0.36409   &    552.77$\pm$51.90    &    4.61$\pm$0.01 &    410.88$\pm$0.23 &    0.23 \\
  G029.5780$-$00.2686$\dagger$  &   29.578     &    -0.26856   &    6.22$\pm$1.00       &    1.50$\pm$0.09 &    1.98$\pm$0.17   &    0.90 \\
  G028.6082$+$00.0185$\dagger$  &   28.6082    &    0.01854    &    210.15$\pm$20.28    &    3.62$\pm$0.01 &    168.17$\pm$0.29 &    0.18 \\
  G029.9559$-$00.0168$\dagger$  &   29.95585   &    -0.01677   &    3116.20$\pm$296.94  &    9.62$\pm$0.01 &    1610.75$\pm$1.84 &    0.52 \\
  G030.0096$-$00.2734    &   30.00965   &    -0.27344   &    4.54$\pm$0.94       &    1.62$\pm$0.14 &    0.3                &    0.77             \\
  G030.5353$+$00.0204$\dagger$  &   30.53526   &    0.02038    &    710.36$\pm$66.36    &    6.30$\pm$0.01 &    553.56$\pm$0.71 &    0.20 \\
  G030.5887$-$00.0428$\dagger$  &   30.58875   &    -0.04278   &    92.37$\pm$8.33      &    1.79$\pm$0.03 &    7.90$\pm$0.48   &    1.93 \\
  G030.7197$-$00.0829$\dagger$  &   30.71968   &    -0.08286   &    969.33$\pm$96.01    &    4.59$\pm$0.01 &    464.58$\pm$2.28 &    0.58 \\
  G030.8662$+$00.1143$\dagger$  &   30.8662    &    0.11429    &    325.47$\pm$32.96    &    3.09$\pm$0.01 &    137.17$\pm$0.60 &    0.68 \\
  G031.0495$+$00.4697    &   31.04949   &    0.46972    &    13.64$\pm$1.49      &    1.78$\pm$0.06 &    10.72$\pm$0.69  &    0.19$\pm$0.09 \\
  G031.2420$-$00.1106$\dagger$  &   31.24202   &    -0.11062   &    296.24$\pm$27.05    &    7.81$\pm$0.01 &    174.42$\pm$5.67 &    0.42 \\
  G031.1596$+$00.0448$\dagger$  &   31.15958   &    0.04475    &    23.83$\pm$2.28      &    1.69$\pm$0.04 &    20.66$\pm$0.32  &    0.11 \\
  G031.2801$+$00.0632$\dagger$  &   31.28008   &    0.06322    &    268.86$\pm$25.67    &    9.33$\pm$0.01 &    144.48$\pm$3.82 &    0.49 \\
  G032.4727$+$00.2036    &   32.47274   &    0.20361    &    97.38$\pm$9.67      &    2.26$\pm$0.01 &    55.96$\pm$0.53  &    0.44$\pm$0.08 \\
  G032.7441$-$00.0755    &   32.74408   &    -0.07553   &    7.93$\pm$1.14       &    1.78$\pm$0.10 &    0.34               &    1.09             \\
  G032.7966$+$00.1909$\dagger$  &   32.79658   &    0.19091    &    3123.37$\pm$281.38  &    10.01$\pm$0.00 &    1698.91$\pm$2.30 &    0.48 \\
  G032.9273$+$00.6060$\dagger$  &   32.92726   &    0.60601    &    285.57$\pm$31.27    &    6.81$\pm$0.01 &    229.50$\pm$1.12 &    0.17 \\
  G032.9906$+$00.0385$\dagger$  &   32.99057   &    0.03852    &    157.76$\pm$18.38    &    15.24$\pm$0.04 &    102.80$\pm$1.07 &    0.34 \\
  G033.1328$-$00.0923$\dagger$  &   33.13277   &    -0.09228   &    378.59$\pm$34.75    &    4.02$\pm$0.00 &    173.43$\pm$1.43 &    0.61 \\
  G033.4163$-$00.0036$\dagger$  &   33.41627   &    -0.00358   &    75.16$\pm$9.16      &    8.78$\pm$0.04 &    57.59$\pm$1.61  &    0.21 \\
       \hline
\end{tabular}
\label{summary_young_hii}
\end{table*}    
\addtocounter{table}{-1}
\begin{table*}
 \centering
 \caption[]{-{ \it continuum \rm Information of 120 young positive spectrum \hii\ regions}}
\scriptsize
  \begin{tabular}{p{3cm}p{2cm}p{2cm}p{2cm}p{2cm}p{2cm}p{2cm}}
  \hline
   \hline
Name	 &	$\ell$	&	$b$	&	Flux$_{5\,GHz}$	&	Angular diameter       &Flux$_{1.4\,GHz}$    &Spectral Index \\
Gal	       &	($\degr$)	&	($\degr$)	&	(mJy) 	&	(\arcsec)	     &	(mJy)	                          &		      \\
\hline
  G033.9145$+$00.1105$\dagger$  &   33.9145    &    0.11045    &    842.22$\pm$88.66    &    10.14$\pm$0.01 &    464.68$\pm$1.90 &    0.47 \\
  G034.2572$+$00.1535$\dagger$  &   34.25724   &    0.15352    &    1762.63$\pm$163.28  &    5.75$\pm$0.01 &    370.78$\pm$3.39 &    1.22 \\
  G034.4032$+$00.2277    &   34.4032    &    0.22771    &    8.92$\pm$1.24       &    1.74$\pm$0.09 &    5.27$\pm$0.51   &    0.41$\pm$0.11 \\
  G035.0242$+$00.3502    &   35.02419   &    0.3502     &    11.44$\pm$1.23      &    1.55$\pm$0.04 &    5.34$\pm$0.40   &    0.60$\pm$0.08 \\
  G035.4669$+$00.1394$\dagger$  &   35.46693   &    0.13942    &    317.60$\pm$29.36    &    5.12$\pm$0.01 &    235.14$\pm$1.11 &    0.24 \\
  G035.5781$-$00.0305$\dagger$    &   35.57813   &    -0.03048   &    187.75$\pm$18.44    &    2.53$\pm$0.01 &    38.05$\pm$0.97  &    1.25 \\
  G036.4057$+$00.0226$\dagger$  &   36.40568   &    0.02256    &    22.34$\pm$2.21      &    1.86$\pm$0.05 &    17.34$\pm$1.12  &    0.20 \\
  G037.5457$-$00.1120$\dagger$  &   37.54572   &    -0.11199   &    406.46$\pm$41.28    &    8.18$\pm$0.01 &    252.32$\pm$1.04 &    0.37 \\
  G037.7347$-$00.1128    &   37.73473   &    -0.11277   &    16.02$\pm$1.63      &    1.76$\pm$0.05 &    12.27$\pm$0.93  &    0.21$\pm$0.08 \\
  G037.7633$-$00.2167$\dagger$  &   37.76327   &    -0.21667   &    337.64$\pm$38.01    &    14.56$\pm$0.03 &    295.32$\pm$2.42 &    0.11 \\
  G037.8731$-$00.3996$\dagger$  &   37.87308   &    -0.39961   &    2561.21$\pm$234.04  &    8.92$\pm$0.00 &    1279.35$\pm$1.16 &    0.55 \\
  G037.9723$-$00.0965$\dagger$  &   37.97232   &    -0.09653   &    20.89$\pm$2.52      &    2.45$\pm$0.12 &    10.22$\pm$0.91  &    0.56 \\
  G038.8756$+$00.3080$\dagger$  &   38.87564   &    0.308      &    311.31$\pm$29.87    &    3.57$\pm$0.01 &    191.23$\pm$1.57 &    0.38 \\
  G039.1956$+$00.2255    &   39.19557   &    0.22546    &    62.27$\pm$6.41      &    1.95$\pm$0.01 &    14.19$\pm$2.24  &    1.16$\pm$0.08 \\
  G039.7277$-$00.3973$\dagger$  &   39.72773   &    -0.39732   &    133.30$\pm$18.10    &    15.94$\pm$0.05 &    112.30$\pm$1.65 &    0.13 \\
  G039.8824$-$00.3460$\dagger$  &   39.88236   &    -0.34602   &    276.87$\pm$26.38    &    3.81$\pm$0.01 &    246.97$\pm$1.45 &    0.09 \\
  G042.4345$-$00.2605$\dagger$  &   42.43453   &    -0.26049   &    83.65$\pm$9.25      &    3.63$\pm$0.02 &    38.47$\pm$2.08  &    0.61 \\
  G043.1651$-$00.0283$\dagger$  &   43.1651    &    -0.02828   &    2714.29$\pm$262.82  &    9.61$\pm$0.01 &    564.34$\pm$8.16 &    1.23 \\
  G043.1665$+$00.0106$\dagger$  &   43.16648   &    0.01057    &    1365.68$\pm$125.16  &    3.68$\pm$0.00 &    237.81$\pm$8.09 &    1.37 \\
  G043.1778$-$00.5181$\dagger$  &   43.17775   &    -0.51806   &    181.65$\pm$23.04    &    7.16$\pm$0.03 &    122.91$\pm$1.30 &    0.31 \\
  G045.0694$+$00.1323$\dagger$  &   45.06939   &    0.1323     &    46.17$\pm$4.44      &    1.96$\pm$0.04 &    17.93$\pm$1.47  &    0.74 \\
  G045.0712$+$00.1321$\dagger$ &  45.07116      &  0.13206    &   146.67$\pm$14.65 &     1.89$\pm$0.01  &     61.6$\pm$0.6 &     0.68  \\ 
  G045.1223$+$00.1321$\dagger$  &   45.12233   &    0.13206    &    2984.27$\pm$274.33  &    7.46$\pm$0.00 &    1345.96$\pm$1.67 &    0.63  \\
  G045.4545$+$00.0591$\dagger$  &   45.4545    &    0.05908    &    1029.45$\pm$98.24   &    7.61$\pm$0.01 &    492.46$\pm$3.21 &    0.58  \\
  G045.4656$+$00.0452    &   45.46557   &    0.04515    &    62.26$\pm$5.79      &    1.70$\pm$0.03 &    28.88$\pm$1.32  &    0.60$\pm$0.07 \\
  G045.4790$+$00.1294$\dagger$  &   45.47896   &    0.12942    &    504.23$\pm$58.87    &    13.72$\pm$0.03 &    380.22$\pm$3.59 &    0.22  \\
  G048.6057$+$00.0228    &   48.60569   &    0.02278    &    36.16$\pm$3.59      &    1.74$\pm$0.04 &    6.61$\pm$1.05   &    1.33$\pm$0.08 \\
  G048.6099$+$00.0270$\dagger$  &   48.60992   &    0.02697    &    131.22$\pm$15.55    &    7.06$\pm$0.03 &    56.49$\pm$1.07  &    0.66  \\
  G048.9296$-$00.2793$\dagger$  &   48.92959   &    -0.27926   &    185.39$\pm$19.22    &    6.16$\pm$0.02 &    66.90$\pm$3.92  &    0.80  \\
  G049.2679$-$00.3374    &   49.26792   &    -0.33743   &    102.61$\pm$14.03    &    6.22$\pm$0.50 &    64.41$\pm$1.51  &    0.37$\pm$0.11 \\
  G049.3704$-$00.3012$\dagger$  &   49.37036   &    -0.30117   &    414.43$\pm$47.36    &    4.68$\pm$0.24 &    252.86$\pm$8.37 &    0.39  \\
  G049.4905$-$00.3688$\dagger$  &   49.49053   &    -0.36881   &    3821.72$\pm$365.24  &    5.76$\pm$0.01 &    1165.61$\pm$5.68 &    0.93  \\
  G050.0457$+$00.7683    &   50.04574   &    0.76833    &    15.57$\pm$1.55      &    2.04$\pm$0.05 &    13.03$\pm$0.39  &    0.14$\pm$0.08 \\
  G050.3152$+$00.6762    &   50.31525   &    0.67623    &    81.31$\pm$8.07      &    2.11$\pm$0.01 &    38.74$\pm$2.01  &    0.58$\pm$0.08 \\
  G050.3157$+$00.6747$\dagger$  &   50.31566   &    0.67475    &    73.26$\pm$8.30      &    9.82$\pm$0.03 &    15.88$\pm$1.99  &    1.20  \\
  G051.6785$+$00.7193    &   51.67854   &    0.71934    &    22.55$\pm$2.07      &    1.79$\pm$0.03 &    2.8                &    0.3              \\
  G052.7533$+$00.3340$\dagger$  &   52.75329   &    0.33397    &    386.03$\pm$36.54    &    8.66$\pm$0.01 &    264.03$\pm$2.23 &    0.30  \\
  G053.9589$+$00.0320$\dagger$  &   53.95892   &    0.03201    &    46.00$\pm$4.73      &    2.31$\pm$0.01 &    40.78$\pm$1.57  &    0.09  \\
  G057.5474$-$00.2717$\dagger$  &   57.54739   &    -0.27167   &    233.68$\pm$22.99    &    13.79$\pm$0.01 &    118.26$\pm$1.38 &    0.54  \\
  G060.8838$-$00.1295$\dagger$  &   60.88377   &    -0.12955   &    292.06$\pm$29.15    &    22.76$\pm$0.03 &    80.89$\pm$1.22  &    1.01  \\
  G061.4763$+$00.0892$\dagger$  &   61.47631   &    0.08921    &    718.71$\pm$64.37    &    6.54$\pm$0.00 &    252.71$\pm$1.40 &    0.82  \\
 \hline  
 \hline 
 \end{tabular}
 \begin{tablenotes}
      \scriptsize
\item These columns contain the name and Galactic coordinate of each source, the flux density and angular diameter of each source at 5\,{\rm GHz} from CORNISH,  flux densities at 1.4\,GHz from THOR, MAGPIS and White2005, as well as  the spectral indices and its errors. Symbol $\dagger$ means that those objects are detected at both 5\,GHz and 1.4\,GHz with lower limit of the spectral indices as they are extended at 1.4 GHz. Flux densities of some sources at 1.4\,GHz with no errors refer to the 1.4\,GHz noise level at the source position, indicating that these sources are only detected at 5\,GHz and so have lower limits of spectral indices. 
\end{tablenotes}
\end{table*}

\begin{table*}
 \caption[]{Information of total 534 positive spectrum radio objects}
\scriptsize
\begin{tabular}{p{3cm}p{2cm}p{2cm}p{2cm}p{2cm}p{2cm}p{2cm}}
  \hline
   \hline
Name	 &	$\ell$	&	$b$	&	Flux$_{5\,GHz}$	&	Angular diameter       &Flux$_{1.4\,GHz}$    &Spectral Index \\
Gal	       &	($\degr$)	&	($\degr$)	&	(mJy) 	&	(\arcsec)	     &	(mJy)	                          &		      \\
\hline
  \hline
  G010.3009$-$00.1477$\dagger$  &   10.30088   &    -0.1477    &    631.39$\pm$59.30    &    5.45$\pm$0.01 &    426.18$\pm$1.98 &    0.31  \\
  G010.3377$+$01.0601    &   10.33767   &    1.06007    &    10.54$\pm$1.22      &    1.50$\pm$0.05 &    1.15               &    0.38             \\
  G010.4168$+$00.9356    &   10.41684   &    0.93556    &    14.38$\pm$1.51      &    1.58$\pm$0.04 &    1.3                &    0.54             \\
  G010.4724$+$00.0275$\dagger$  &   10.47236   &    0.0275     &    38.43$\pm$4.38      &    2.24$\pm$0.02 &    31.26$\pm$0.35  &    0.16  \\
  G010.4727$-$00.6742$\dagger$  &   10.47275   &    -0.67423   &    6.28$\pm$1.18       &    1.90$\pm$0.16 &    2.84$\pm$0.25   &    0.62  \\
  G010.6223$-$00.3788$\dagger$  &   10.62231   &    -0.37877   &    483.33$\pm$49.87    &    5.76$\pm$0.01 &    327.63$\pm$1.89 &    0.31  \\
\hline
\end{tabular}
\label{summary_positive_objects}
\end{table*}    
\addtocounter{table}{-1}
\begin{table*}
 \caption[]{-continuum {Information of total 534 positive spectrum radio objects}}
\scriptsize
  \begin{tabular}{p{3cm}p{2cm}p{2cm}p{2cm}p{2cm}p{2cm}p{2cm}}
  \hline
   \hline
Name	 &	$\ell$	&	$b$	&	Flux$_{5\,GHz}$	&	Angular diameter       &Flux$_{1.4\,GHz}$    &Spectral Index \\
Gal &	($\degr$)	&	($\degr$)	&	(mJy) 	&	(\arcsec) &	(mJy)	                          & \\
\hline
  \hline
  G017.8409$-$00.1765$\dagger$  &   17.8409    &    -0.17647   &    6.58$\pm$0.82       &    1.64$\pm$0.07 &    3.77$\pm$0.21   &    0.44  \\
  G017.8645$+$00.2120$\dagger$  &   17.86448   &    0.21199    &    23.35$\pm$2.16      &    1.57$\pm$0.03 &    6.97$\pm$0.20   &    0.95  \\
  G017.9275$+$00.6337    &   17.92751   &    0.63366    &    4.40$\pm$0.71       &    1.50$\pm$0.09 &    0.1                &    1.56             \\
  G018.0661$+$00.8535    &   18.0661    &    0.85354    &    33.16$\pm$3.24      &    2.83$\pm$0.08 &    20.09$\pm$1.33  &    0.39$\pm$0.08 \\
  G018.1286$-$00.2189$\dagger$  &   18.12862   &    -0.2189    &    8.54$\pm$1.70       &    1.95$\pm$0.18 &    5.17$\pm$0.36   &    0.39  \\
  G018.1460$-$00.2839$\dagger$  &   18.14602   &    -0.2839    &    856.18$\pm$82.85    &    23.42$\pm$0.02 &    151.22$\pm$1.93 &    1.36  \\
\hline
\end{tabular}
\label{summary_positive_objects}
\end{table*}    
\addtocounter{table}{-1}
\begin{table*}
 \caption[]{-continuum {Information of total 534 positive spectrum radio objects}}
\scriptsize
  \begin{tabular}{p{3cm}p{2cm}p{2cm}p{2cm}p{2cm}p{2cm}p{2cm}}
  \hline
   \hline
Name	 &	$\ell$	&	$b$	&	Flux$_{5\,GHz}$	&	Angular diameter       &Flux$_{1.4\,GHz}$    &Spectral Index \\
Gal	       &	($\degr$)	&	($\degr$)	&	(mJy) 	&	(\arcsec)	     &	(mJy)	                          &		      \\
\hline
  \hline
  G021.9836$+$00.7962    &   21.98358   &    0.79623    &    5.53$\pm$0.77       &    1.50$\pm$0.07 &    4.56$\pm$0.18   &    0.15$\pm$0.11 \\
  G021.9972$-$00.8838    &   21.99717   &    -0.88375   &    14.22$\pm$1.65      &    2.03$\pm$0.08 &    10.83$\pm$0.38  &    0.21$\pm$0.09 \\
  G022.1540$-$00.1542$\dagger$  &   22.15397   &    -0.15417   &    22.29$\pm$2.10      &    1.70$\pm$0.03 &    16.34$\pm$0.18  &    0.24  \\
  G022.1794$+$00.7537    &   22.17939   &    0.75374    &    39.08$\pm$3.51      &    1.51$\pm$0.02 &    28.17$\pm$0.24  &    0.26$\pm$0.07 \\
  G022.2211$+$00.9009    &   22.22111   &    0.90093    &    21.38$\pm$2.39      &    2.46$\pm$0.10 &    14.31$\pm$0.77  &    0.32$\pm$0.09 \\
  G022.5477$-$00.1061$\dagger$  &   22.54772   &    -0.10612   &    32.09$\pm$3.03      &    2.01$\pm$0.04 &    24.94$\pm$1.05  &    0.20 \\
\hline
\end{tabular}
\label{summary_positive_objects}
\end{table*}    
\addtocounter{table}{-1}
\begin{table*}
 \caption[]{-continuum {Information of total 534 positive spectrum radio objects}}
\scriptsize
  \begin{tabular}{p{3cm}p{2cm}p{2cm}p{2cm}p{2cm}p{2cm}p{2cm}}
  \hline
   \hline
Name	 &	$\ell$	&	$b$	&	Flux$_{5\,GHz}$	&	Angular diameter       &Flux$_{1.4\,GHz}$    &Spectral Index \\
Gal	       &	($\degr$)	&	($\degr$)	&	(mJy) 	&	(\arcsec)	     &	(mJy) &		      \\
\hline
  \hline
  G029.1382$+$00.8332    &   29.1382    &    0.83317    &    13.74$\pm$1.46      &    1.50$\pm$0.04 &    6.36$\pm$0.23   &    0.61$\pm$0.08 \\
  G029.2555$-$00.8653    &   29.25545   &    -0.86532   &    3.16$\pm$0.81       &    1.73$\pm$0.20 &    0.39               &    0.17             \\
  G029.2620$+$00.2916    &   29.26197   &    0.29162    &    2.76$\pm$0.75       &    1.50$\pm$0.17 &    0.13               &    0.86             \\
  G029.3096$+$00.5124    &   29.30957   &    0.51242    &    2.84$\pm$0.70       &    1.50$\pm$0.15 &    0.12               &    1.01             \\
  G029.4404$-$00.3199    &   29.44038   &    -0.31986   &    2.81$\pm$0.81       &    1.50$\pm$0.18 &    0.23               &    0.45             \\
  G029.5069$-$01.1293    &   29.50689   &    -1.12928   &    4.22$\pm$1.16       &    1.60$\pm$0.19 &    0.25               &    0.79             \\
\hline
\end{tabular}
\label{summary_positive_objects}
\end{table*}    
\addtocounter{table}{-1}
\begin{table*}
 \caption[]{-continuum {Information of total 534 positive spectrum radio objects}}
\scriptsize
  \begin{tabular}{p{3cm}p{2cm}p{2cm}p{2cm}p{2cm}p{2cm}p{2cm}}
  \hline
   \hline
Name	 &	$\ell$	&	$b$	&	Flux$_{5\,GHz}$	&	Angular diameter       &Flux$_{1.4\,GHz}$    &Spectral Index \\
Gal &	($\degr$)	&	($\degr$)	&	(mJy) 	&	(\arcsec) &	(mJy) & \\
\hline
  \hline
  G034.1782$+$00.2564    &   34.17819   &    0.25638    &    8.02$\pm$1.06       &    1.50$\pm$0.06 &    2.82$\pm$0.31   &    0.82$\pm$0.10 \\
  G034.2171$-$00.6886    &   34.21709   &    -0.68863   &    2.37$\pm$0.60       &    1.57$\pm$0.17 &    0.23               &    0.36             \\
  G034.2541$+$00.3982    &   34.25414   &    0.39817    &    7.98$\pm$1.13       &    1.64$\pm$0.08 &    6.06$\pm$0.43   &    0.22$\pm$0.11 \\
  G034.2572$+$00.1535$\dagger$  &   34.25724   &    0.15352    &    1762.63$\pm$163.28  &    5.75$\pm$0.01 &    370.78$\pm$3.39 &    1.22  \\
  G034.2655$+$00.7195    &   34.26547   &    0.71948    &    6.96$\pm$1.10       &    1.76$\pm$0.11 &    2.38$\pm$0.28   &    0.84$\pm$0.12 \\
  G034.2830$+$00.0087    &   34.28298   &    0.00872    &    7.98$\pm$1.62       &    2.31$\pm$0.24 &    0.2                &    1.63             \\
\hline
\end{tabular}
\label{summary_positive_objects}
\end{table*}    
\addtocounter{table}{-1}
\begin{table*}
 \caption[]{-continuum {Information of total 534 positive spectrum radio objects}}
\scriptsize
  \begin{tabular}{p{3cm}p{2cm}p{2cm}p{2cm}p{2cm}p{2cm}p{2cm}}
  \hline
   \hline
Name	 &	$\ell$	&	$b$	&	Flux$_{5\,GHz}$	&	Angular diameter       &Flux$_{1.4\,GHz}$    &Spectral Index \\
Gal &	($\degr$)	&	($\degr$)	&	(mJy) 	&	(\arcsec) &	(mJy) &		      \\
\hline
  \hline
  G043.1665$+$00.0106$\dagger$  &   43.16648   &    0.01057    &    1365.68$\pm$125.16  &    3.68$\pm$0.00 &    237.81$\pm$8.09 &    1.37  \\
  G043.1778$-$00.5181$\dagger$  &   43.17775   &    -0.51806   &    181.65$\pm$23.04    &    7.16$\pm$0.03 &    122.91$\pm$1.30 &    0.31  \\
  G043.1845$-$00.5268$\dagger$  &   43.1845    &    -0.52684   &    352.51$\pm$37.69    &    18.96$\pm$0.03 &    306.10$\pm$1.33 &    0.11  \\
  G043.2946$-$00.6455$\dagger$  &   43.29464   &    -0.64553   &    68.05$\pm$6.11      &    1.81$\pm$0.03 &    26.64$\pm$1.80  &    0.74  \\
  G043.5793$+$00.0261    &   43.57932   &    0.02608    &    16.00$\pm$1.52      &    1.55$\pm$0.03 &    8.94$\pm$0.35   &    0.46$\pm$0.07 \\
  G043.8056$-$00.7210    &   43.80561   &    -0.72096   &    13.37$\pm$1.33      &    1.59$\pm$0.04 &    10.94$\pm$2.80  &    0.16$\pm$0.08 \\
\hline
\end{tabular}
\label{summary_positive_objects}
\end{table*}    
\addtocounter{table}{-1}
\begin{table*}
 \caption[]{-continuum {Information of total 534 positive spectrum radio objects}}
\scriptsize
  \begin{tabular}{p{3cm}p{2cm}p{2cm}p{2cm}p{2cm}p{2cm}p{2cm}}
  \hline
   \hline
Name	 &	$\ell$	&	$b$	&	Flux$_{5\,GHz}$	&	Angular diameter       &Flux$_{1.4\,GHz}$    &Spectral Index \\
Gal &	($\degr$)	&	($\degr$)	&	(mJy) &	(\arcsec) &	(mJy)	                          & \\
\hline
  \hline
  G051.5095$+$00.1686$\dagger$  &   51.50955   &    0.16864    &    153.38$\pm$14.17    &    7.51$\pm$0.01 &    43.27$\pm$1.15  &    0.99  \\
  G051.5869$+$00.4807    &   51.58695   &    0.48073    &    9.22$\pm$0.93       &    1.50$\pm$0.04 &    0.98               &    0.41             \\
  G051.6785$+$00.7193    &   51.67854   &    0.71934    &    22.55$\pm$2.07      &    1.79$\pm$0.03 &    2.8                &    0.3              \\
  G051.7134$-$00.0192    &   51.71341   &    -0.01917   &    9.13$\pm$0.91       &    1.50$\pm$0.04 &    8.13$\pm$1.31   &    0.09$\pm$0.08 \\
  G051.8062$+$00.0389    &   51.80622   &    0.03893    &    9.28$\pm$0.92       &    1.50$\pm$0.04 &    1.31               &    0.19             \\
  G051.8341$+$00.2838    &   51.83413   &    0.28378    &    76.00$\pm$6.81      &    2.25$\pm$0.03 &    54.55$\pm$2.00  &    0.26$\pm$0.07 \\
\hline
\hline
\end{tabular}
\begin{tablenotes}
      \scriptsize
\item These columns contain the name and Galactic coordinate of each source, the flux density and angular diameter of each source at 5\,{\rm GHz} from CORNISH,  flux densities at 1.4\,GHz from THOR, MAGPIS and White2005, as well as  the spectral indices and its errors. Symbol $\dagger$ means that those objects are detected at both 5\,GHz and 1.4\,GHz with lower limit of the spectral indices as they are extended at 1.4 GHz. Flux densities of some sources at 1.4\,GHz with no errors refer to the noise level at 1.4\,GHz to the source position, indicating that these sources are only detected at 5\,GHz with the lower limits of spectral indices. 
\end{tablenotes}
\label{summary_positive_objects}
\end{table*} 


\bibliographystyle{mn2e}
\bibliography{ref}

\begin{thebibliography}{}

\bibitem[\protect\citeauthoryear{{Afflerbach}, {Churchwell}, {Acord}, {Hofner},
  {Kurtz} \& {Depree}}{{Afflerbach} et~al.}{1996}]{Afflerbach1996ApJS106}
{Afflerbach} A.,  {Churchwell} E.,  {Acord} J.~M.,  {Hofner} P.,  {Kurtz} S.,
   {Depree} C.~G.,  1996, \apjs, 106, 423

\bibitem[\protect\citeauthoryear{{Avalos}, {Lizano}, {Rodr{\'{\i}}guez},
  {Franco-Hern{\'a}ndez} \& {Moran}}{{Avalos} et~al.}{2006}]{Avalos2006ApJ}
{Avalos} M.,  {Lizano} S.,  {Rodr{\'{\i}}guez} L.~F.,  {Franco-Hern{\'a}ndez}
  R.,    {Moran} J.~M.,  2006, \apj, 641, 406

\bibitem[\protect\citeauthoryear{{Beltr{\'a}n}, {Cesaroni}, {Moscadelli} \&
  {Codella}}{{Beltr{\'a}n} et~al.}{2007}]{Beltran2007AA}
{Beltr{\'a}n} M.~T.,  {Cesaroni} R.,  {Moscadelli} L.,    {Codella} C.,  2007,
  \aap, 471, L13

\bibitem[\protect\citeauthoryear{{Beuther} et~al.,}{{Beuther}
  et~al.}{2016}]{Beuther2016AA595A32B}
{Beuther} H.  et~al., 2016, \aap, 595, A32

\bibitem[\protect\citeauthoryear{{Beuther}, {Churchwell}, {McKee} \&
  {Tan}}{{Beuther} et~al.}{2007}]{Beuther2007prpl165B}
{Beuther} H.,  {Churchwell} E.~B.,  {McKee} C.~F.,    {Tan} J.~C.,  2007,
  Protostars and Planets V, pp 165--180

\bibitem[\protect\citeauthoryear{{Bihr} et~al.,}{{Bihr}
  et~al.}{2016}]{Bihr2016AA}
{Bihr} S.  et~al., 2016, \aap, 588, A97

\bibitem[\protect\citeauthoryear{{Cesaroni} et~al.,}{{Cesaroni}
  et~al.}{2015}]{Cesaroni2015AA579A}
{Cesaroni} R.  et~al., 2015, \aap, 579, A71

\bibitem[\protect\citeauthoryear{{Davies}, {Hoare}, {Lumsden}, {Hosokawa},
  {Oudmaijer}, {Urquhart}, {Mottram} \& {Stead}}{{Davies}
  et~al.}{2011}]{Davies2011MNRAS972D}
{Davies} B.,  {Hoare} M.~G.,  {Lumsden} S.~L.,  {Hosokawa} T.,  {Oudmaijer}
  R.~D.,  {Urquhart} J.~S.,  {Mottram} J.~C.,    {Stead} J.,  2011, \mnras,
  416, 972

\bibitem[\protect\citeauthoryear{{de Pree}, {Gaume}, {Goss} \& {Claussen}}{{de
  Pree} et~al.}{1996}]{dePree1996ApJ}
{de Pree} C.~G.,  {Gaume} R.~A.,  {Goss} W.~M.,    {Claussen} M.~J.,  1996,
  \apj, 464, 788

\bibitem[\protect\citeauthoryear{{De Pree}, {Mehringer} \& {Goss}}{{De Pree}
  et~al.}{1997}]{dePree1997ApJ}
{De Pree} C.~G.,  {Mehringer} D.~M.,    {Goss} W.~M.,  1997, \apj, 482, 307

\bibitem[\protect\citeauthoryear{{De Pree}, {Wilner}, {Mercer}, {Davis}, {Goss}
  \& {Kurtz}}{{De Pree} et~al.}{2004}]{DePree2004ApJ}
{De Pree} C.~G.,  {Wilner} D.~J.,  {Mercer} A.~J.,  {Davis} L.~E.,  {Goss}
  W.~M.,    {Kurtz} S.,  2004, \apj, 600, 286

\bibitem[\protect\citeauthoryear{{Dyson}, {Williams} \& {Redman}}{{Dyson}
  et~al.}{1995}]{Dyson1995MNRAS277}
{Dyson} J.~E.,  {Williams} R.~J.~R.,    {Redman} M.~P.,  1995, \mnras, 277, 700

\bibitem[\protect\citeauthoryear{{Ellingsen}, {Shabala} \& {Kurtz}}{{Ellingsen}
  et~al.}{2005}]{Ellingsen2005MNRAS357}
{Ellingsen} S.~P.,  {Shabala} S.~S.,    {Kurtz} S.~E.,  2005, \mnras, 357, 1003

\bibitem[\protect\citeauthoryear{{Gaume}, {Goss}, {Dickel}, {Wilson} \&
  {Johnston}}{{Gaume} et~al.}{1995}]{Gaume1995ApJ}
{Gaume} R.~A.,  {Goss} W.~M.,  {Dickel} H.~R.,  {Wilson} T.~L.,    {Johnston}
  K.~J.,  1995, \apj, 438, 776

\bibitem[\protect\citeauthoryear{{Hancock}, {Murphy}, {Gaensler}, {Hopkins} \&
  {Curran}}{{Hancock} et~al.}{2012}]{Hancock2012MNRAS}
{Hancock} P.~J.,  {Murphy} T.,  {Gaensler} B.~M.,  {Hopkins} A.,    {Curran}
  J.~R.,  2012, \mnras, 422, 1812

\bibitem[\protect\citeauthoryear{{Helfand}, {Becker}, {White}, {Fallon} \&
  {Tuttle}}{{Helfand} et~al.}{2006}]{Helfand2006AJ}
{Helfand} D.~J.,  {Becker} R.~H.,  {White} R.~L.,  {Fallon} A.,    {Tuttle} S.,
   2006, \aj, 131, 2525

\bibitem[\protect\citeauthoryear{{Hoare}, {Kurtz}, {Lizano}, {Keto} \&
  {Hofner}}{{Hoare} et~al.}{2007}]{Hoare2007prplconfH}
{Hoare} M.~G.,  {Kurtz} S.~E.,  {Lizano} S.,  {Keto} E.,    {Hofner} P.,  2007,
  Protostars and Planets V, pp 181--196

\bibitem[\protect\citeauthoryear{{Hoare} et~al.,}{{Hoare}
  et~al.}{2012}]{Hoare2012PASP}
{Hoare} M.~G.  et~al., 2012, \pasp, 124, 939

\bibitem[\protect\citeauthoryear{{Hosokawa} \& {Omukai}}{{Hosokawa} \&
  {Omukai}}{2009}]{Hosokawa2009ApJ691}
{Hosokawa} T.,  {Omukai} K.,  2009, \apj, 691, 823

\bibitem[\protect\citeauthoryear{{Hosokawa}, {Yorke} \& {Omukai}}{{Hosokawa}
  et~al.}{2010}]{Hosokawa2010ApJ721}
{Hosokawa} T.,  {Yorke} H.~W.,    {Omukai} K.,  2010, \apj, 721, 478

\bibitem[\protect\citeauthoryear{{Johnson}, {De Pree} \& {Goss}}{{Johnson}
  et~al.}{1998}]{Johnson1998ApJ}
{Johnson} C.~O.,  {De Pree} C.~G.,    {Goss} W.~M.,  1998, \apj, 500, 302

\bibitem[\protect\citeauthoryear{{Keto}}{{Keto}}{2003}]{Keto2003ApJ}
{Keto} E.,  2003, \apj, 599, 1196

\bibitem[\protect\citeauthoryear{{Keto}, {Zhang} \& {Kurtz}}{{Keto}
  et~al.}{2008}]{Keto2008ApJ672}
{Keto} E.,  {Zhang} Q.,    {Kurtz} S.,  2008, \apj, 672, 423

\bibitem[\protect\citeauthoryear{{Kim} \& {Koo}}{{Kim} \&
  {Koo}}{2001}]{Kim2001ApJ549}
{Kim} K.-T.,  {Koo} B.-C.,  2001, \apj, 549, 979

\bibitem[\protect\citeauthoryear{{Kim}, {Wyrowski}, {Urquhart}, {Menten} \&
  {Csengeri}}{{Kim} et~al.}{2017}]{Kim2017AA602A}
{Kim} W.-J.,  {Wyrowski} F.,  {Urquhart} J.~S.,  {Menten} K.~M.,    {Csengeri}
  T.,  2017, \aap, 602, A37

\bibitem[\protect\citeauthoryear{{Kurtz}}{{Kurtz}}{2005}]{Kurtz2005IAUS}
{Kurtz} S.,  2005, in {Cesaroni} R.,  {Felli} M.,  {Churchwell} E.,
  {Walmsley} M.,  eds,  IAU Symposium Vol. 227, Massive Star Birth: A
  Crossroads of Astrophysics. pp 111--119

\bibitem[\protect\citeauthoryear{{Kurtz}, {Churchwell} \& {Wood}}{{Kurtz}
  et~al.}{1994}]{Kurtz1994ApJS659K}
{Kurtz} S.,  {Churchwell} E.,    {Wood} D.~O.~S.,  1994, \apjs, 91, 659

\bibitem[\protect\citeauthoryear{{Lumsden}, {Hoare}, {Urquhart}, {Oudmaijer},
  {Davies}, {Mottram}, {Cooper} \& {Moore}}{{Lumsden}
  et~al.}{2013}]{Lumsden2013ApJS208}
{Lumsden} S.~L.,  {Hoare} M.~G.,  {Urquhart} J.~S.,  {Oudmaijer} R.~D.,
  {Davies} B.,  {Mottram} J.~C.,  {Cooper} H.~D.~B.,    {Moore} T.~J.~T.,
  2013, \apjs, 208, 11

\bibitem[\protect\citeauthoryear{{McKee} \& {Tan}}{{McKee} \&
  {Tan}}{2003}]{McKee2003ApJM}
{McKee} C.~F.,  {Tan} J.~C.,  2003, \apj, 585, 850

\bibitem[\protect\citeauthoryear{{Molinari} et~al.,}{{Molinari}
  et~al.}{2016}]{Molinari2016AA591A}
{Molinari} S.  et~al., 2016, \aap, 591, A149

\bibitem[\protect\citeauthoryear{{Molinari} et~al.,}{{Molinari}
  et~al.}{2010}]{Molinari2010PASP}
{Molinari} S.  et~al., 2010, \pasp, 122, 314

\bibitem[\protect\citeauthoryear{{Murphy}, {Cohen}, {Ekers}, {Green}, {Wark} \&
  {Moss}}{{Murphy} et~al.}{2010}]{Murphy2010MNRASa2}
{Murphy} T.,  {Cohen} M.,  {Ekers} R.~D.,  {Green} A.~J.,  {Wark} R.~M.,
  {Moss} V.,  2010, \mnras, 405, 1560

\bibitem[\protect\citeauthoryear{{Peters}, {Banerjee}, {Klessen}, {Mac Low},
  {Galv{\'a}n-Madrid} \& {Keto}}{{Peters} et~al.}{2010}]{Peters2010ApJ711}
{Peters} T.,  {Banerjee} R.,  {Klessen} R.~S.,  {Mac Low} M.-M.,
  {Galv{\'a}n-Madrid} R.,    {Keto} E.~R.,  2010, \apj, 711, 1017

\bibitem[\protect\citeauthoryear{{Purcell} et~al.,}{{Purcell}
  et~al.}{2013}]{Purcell2013ApJS}
{Purcell} C.~R.  et~al., 2013, \apjs, 205, 1

\bibitem[\protect\citeauthoryear{{Schuller} et~al.,}{{Schuller}
  et~al.}{2009}]{Schuller2009AA}
{Schuller} F.  et~al., 2009, \aap, 504, 415

\bibitem[\protect\citeauthoryear{{Sewilo}, {Churchwell}, {Kurtz}, {Goss} \&
  {Hofner}}{{Sewilo} et~al.}{2004}]{Sewilo2004ApJ}
{Sewilo} M.,  {Churchwell} E.,  {Kurtz} S.,  {Goss} W.~M.,    {Hofner} P.,
  2004, \apj, 605, 285

\bibitem[\protect\citeauthoryear{{Sewi{\l}o}, {Churchwell}, {Kurtz}, {Goss} \&
  {Hofner}}{{Sewi{\l}o} et~al.}{2011}]{Sewilo2011ApJS}
{Sewi{\l}o} M.,  {Churchwell} E.,  {Kurtz} S.,  {Goss} W.~M.,    {Hofner} P.,
  2011, \apjs, 194, 44

\bibitem[\protect\citeauthoryear{{Shepherd}, {Churchwell} \& {Goss}}{{Shepherd}
  et~al.}{1995}]{Shepherd1995ApJ}
{Shepherd} D.~S.,  {Churchwell} E.,    {Goss} W.~M.,  1995, \apj, 448, 426

\bibitem[\protect\citeauthoryear{{Shi}, {Zhao} \& {Han}}{{Shi}
  et~al.}{2010}]{Shi2010ApJ843S}
{Shi} H.,  {Zhao} J.-H.,    {Han} J.~L.,  2010, \apj, 710, 843

\bibitem[\protect\citeauthoryear{{Tan} \& {McKee}}{{Tan} \&
  {McKee}}{2003}]{Tan20039139T}
{Tan} J.~C.,  {McKee} C.~F.,  2003, ArXiv Astrophysics e-prints

\bibitem[\protect\citeauthoryear{{Tanaka}, {Tan} \& {Zhang}}{{Tanaka}
  et~al.}{2016}]{Tanaka2016ApJ52T}
{Tanaka} K.~E.~I.,  {Tan} J.~C.,    {Zhang} Y.,  2016, \apj, 818, 52

\bibitem[\protect\citeauthoryear{{Tenorio-Tagle}}{{Tenorio-Tagle}}{1979}]{Tenorio1979AA71}
{Tenorio-Tagle} G.,  1979, \aap, 71, 59

\bibitem[\protect\citeauthoryear{{Thompson} et~al.,}{{Thompson}
  et~al.}{2015}]{thompson2015}
{Thompson} M.  et~al., 2015, Advancing Astrophysics with the Square Kilometre
  Array (AASKA14), p.~126

\bibitem[\protect\citeauthoryear{{Thompson}, {Hatchell}, {Walsh}, {MacDonald}
  \& {Millar}}{{Thompson} et~al.}{2006}]{Thompson2006AA}
{Thompson} M.~A.,  {Hatchell} J.,  {Walsh} A.~J.,  {MacDonald} G.~H.,
  {Millar} T.~J.,  2006, \aap, 453, 1003

\bibitem[\protect\citeauthoryear{{Tian} \& {Leahy}}{{Tian} \&
  {Leahy}}{2005}]{Tian2005AAT}
{Tian} W.~W.,  {Leahy} D.,  2005, \aap, 436, 187

\bibitem[\protect\citeauthoryear{{Tian} \& {Leahy}}{{Tian} \&
  {Leahy}}{2006}]{Tian2006AAT}
{Tian} W.~W.,  {Leahy} D.~A.,  2006, \aap, 447, 205

\bibitem[\protect\citeauthoryear{{Urquhart} et~al.,}{{Urquhart}
  et~al.}{2013}]{Urquhart2013MNRAS}
{Urquhart} J.~S.  et~al., 2013, \mnras, 435, 400

\bibitem[\protect\citeauthoryear{{White}, {Becker} \& {Helfand}}{{White}
  et~al.}{2005}]{White2005AJ}
{White} R.~L.,  {Becker} R.~H.,    {Helfand} D.~J.,  2005, \aj, 130, 586

\bibitem[\protect\citeauthoryear{{Wood} \& {Churchwell}}{{Wood} \&
  {Churchwell}}{1989}]{Wood1989ApJS}
{Wood} D.~O.~S.,  {Churchwell} E.,  1989, \apjs, 69, 831

\bibitem[\protect\citeauthoryear{{Zhang}, {Wang}, {Xu}, {Wyrowski} \&
  {Menten}}{{Zhang} et~al.}{2014}]{Zhang2014ApJ}
{Zhang} C.-P.,  {Wang} J.-J.,  {Xu} J.-L.,  {Wyrowski} F.,    {Menten} K.~M.,
  2014, \apj, 784, 107

\bibitem[\protect\citeauthoryear{{Zinnecker} \& {Yorke}}{{Zinnecker} \&
  {Yorke}}{2007}]{Zinnecker2007ARAA45}
{Zinnecker} H.,  {Yorke} H.~W.,  2007, \araa, 45, 481

\end{thebibliography}

\bsp	
\label{lastpage}
\end{document}